\documentclass[lettersize,journal]{IEEEtran}
\usepackage{amsmath,amsfonts,amsthm}

\usepackage{graphicx}
\usepackage{textcomp}
\usepackage{enumerate}
\usepackage{wrapfig}
\usepackage{float}
\usepackage{diagbox} 

\usepackage{comment}
\usepackage{comment}
\usepackage{listings,color}
\usepackage{multirow}
\usepackage{balance}
\usepackage{url}
\usepackage{array,color,colortbl,graphicx,multirow}
\usepackage{subcaption}
\usepackage{caption}
\usepackage{makecell}
\usepackage{url}
\usepackage{nicematrix}
\usepackage{mathtools}
\usepackage{tikz}
\usetikzlibrary{shapes}
\usetikzlibrary{decorations.markings,shadows, shapes}
\usepackage[linesnumbered,ruled,vlined]{algorithm2e}
\SetKwInput{KwData}{Input}
\SetKwInput{KwResult}{Output}
\SetKwComment{Comment}{/* }{ */}
\usepackage{wrapfig}
\usepackage{enumitem}
\usepackage{booktabs}
\usepackage{marvosym}

\usepackage{etoolbox}
\usepackage{tikz}
\usetikzlibrary{patterns, positioning,shapes.misc,matrix, arrows.meta, calc} %
\usepackage[most]{tcolorbox}  
\usepackage{xcolor} 
\usepackage{siunitx}
\usepackage{tabularx}
\usepackage{amssymb}

\newtheorem{theorem}{Theorem}

\usepackage{etoolbox}
\usepackage[pdfpagelabels=true,hidelinks]{hyperref}

\usepackage[alignedforall]{lpform}
\usepackage{epigraph}

\hypersetup{
    colorlinks=true,
    citecolor=blue,
    linkcolor=blue,
}

\hypersetup{
    urlcolor=black
}

\begin{document}

\title{\huge RailS: Load Balancing for All-to-All Communication in Distributed Mixture-of-Experts Training}

\author{
    \IEEEauthorblockN{Heng Xu$^1$, Zhiwei Yu$^2$, Chengze Du$^{1}$, Ying Zhou$^1$, Letian Li$^3$, Haojie Wang$^4$, Weiqiang Cheng$^4$, Jialong Li$^{1}$\textsuperscript{\Letter}} \\
    \IEEEauthorblockA{\normalsize \textsuperscript{1} Faculty of Computer Science and Control Engineering, Shenzhen University of Advanced Technology\\
    \textsuperscript{2} Institute for Network Sciences and Cyberspace, Tsinghua University\\
    \textsuperscript{3}Department of Information Engineering, Chinese University of Hong Kong\\
    \textsuperscript{4}China Mobile\\
    }
    \IEEEauthorblockA{\textsuperscript{\Letter}\text{\texttt{lijialong@suat-sz.edu.cn} 
    }}
}
\markboth{IEEE/ACM Transactions on Networking,~Vol.~XX, No.~XX, OCTOBER~2025}%
{Shell \MakeLowercase{\textit{et al.}}: A Sample Article Using IEEEtran.cls for IEEE Journals}

\maketitle

\begin{abstract}
Training Mixture-of-Experts (MoE) models introduces sparse and highly imbalanced all-to-all communication that dominates iteration time. Conventional load-balancing methods fail to exploit the deterministic topology of Rail architectures, leaving multi-NIC bandwidth underutilized. We present RailS, a distributed load-balancing framework that minimizes all-to-all completion time in MoE training. RailS leverages the Rail topology’s symmetry to prove that uniform sending ensures uniform receiving, transforming global coordination into local scheduling. Each node independently executes a Longest Processing Time First (LPT) spraying scheduler to proactively balance traffic using local information. RailS activates N parallel rails for fine-grained, topology-aware multipath transmission. Across synthetic and real-world MoE workloads, RailS improves bus bandwidth by 20\%–78\% and reduces completion time by 17\%–78\%. For Mixtral workloads, it shortens iteration time by 18\%–40\% and achieves near-optimal load balance, fully exploiting architectural parallelism in distributed training.
\end{abstract}


\section{Introduction}
\IEEEPARstart{A}{rtificial} intelligence (AI) training has shifted from dense models to mixture-of-experts (MoE) models, making cross-GPU communication a critical bottleneck. While early dense models~\cite{resnet,bert} used structured collective communication over millions of parameters, modern large language models (LLMs)~\cite{gpt3,llama,qwen3} rely on multi-level parallelism and collective operations at the scale of billions of parameters. MoE models~\cite{gshard,switchtransformer,glam,deepspeedmoe,deepseekv3,mixtral,deepseekr1} introduce expert parallelism with tens of billions of parameters, where all-to-all communication generates sparse and highly imbalanced traffic across hundreds of GPUs.

This all-to-all communication makes load balancing a central challenge. Studies show that such communication dominates MoE iterations, with latency limiting distributed training throughput~\cite{mixnet}. To mitigate this, the Rail architecture~\cite{nvidiarail} employs multi-NIC direct-to-leaf links to relieve bandwidth bottlenecks in conventional spine–leaf networks. Nevertheless, existing deployments remain suboptimal for MoE’s irregular traffic patterns and dynamic load-balancing needs, leaving hardware potential largely untapped.

\textbf{Challenge 1: The inherent path selection conflicts in the Rail architecture can completely negate its hardware parallelism advantages.}

Path selection~\cite{yen1971finding,power2,raiciu2011improving,proteus1,proteus2,uccl_transport} critically affects communication efficiency. Rail offers two deterministic path options for all-to-all traffic, each with limitations. The first routes traffic through the spine layer. It works with traditional topologies but relies on hashing~\cite{ecmp,wcmp}, which ignores MoE traffic sparsity. This causes large flows to cluster on a few links, creating congestion while other paths remain idle. The second option uses NIC-direct rails. Without advanced transport support such as multipath transmission, flow splitting, reordering, and reassembly, the bottleneck stays at a single node. Spine paths fail due to architecture limits, and direct rails are underutilized because of immature transport protocols, restricting Rail's parallelism.

\textbf{Challenge 2: Existing load balancing methods overlook topological characteristics, preventing their scheduling decisions from approaching the theoretical upper bound.}

Current load balancing approaches ignore Rail’s deterministic topology. ECMP and dynamic load balancers~\cite{conga,hula,presto,plb,laps} treat multiple paths as interchangeable and select them at random. However, Rail has $N$ parallel rails with a one-to-one mapping from each source NIC to a destination NIC. This predictable and symmetric structure offers prior knowledge useful for global optimization. Existing methods overlook it, so both static hashing and dynamic probing choose paths blindly, fail to exploit the topology, and cannot achieve the theoretical optimum, limiting system throughput.

\begin{figure*}[th]
    \centering
    \includegraphics[width=1\textwidth]{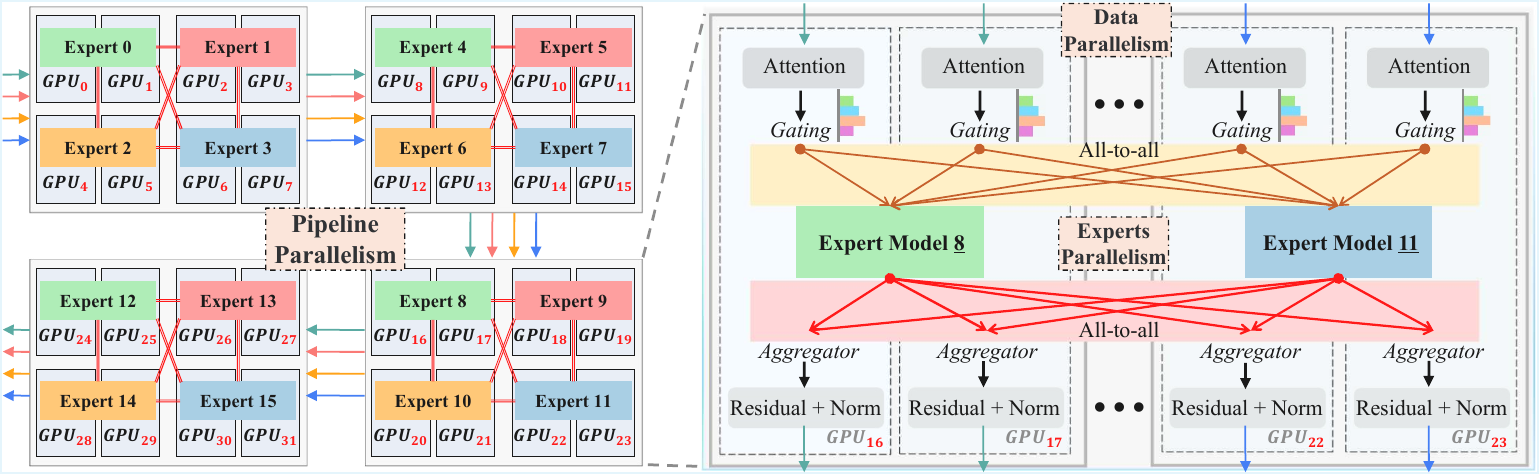}
    \caption{Hybrid parallelism for a MoE model (data parallelism, pipeline parallelism, expert parallelism, and tensor parallelism).}
    \label{fig:moe}
\end{figure*}

\textbf{Challenge 3: The granularity of traffic splitting involves an inherent trade-off, creating a dilemma between load balancing and system efficiency.}

In load balancing, loss is inevitable due to the granularity of traffic splitting. Splitting traffic at the subflow level~\cite{sen2013scalable,minrtt} can lead to collisions and uneven utilization across $N$ NIC paths, even when flows are further divided or recombined. Splitting traffic at the flowlet~\cite{letflow,expeditus,flowbender} or packet level~\cite{dixit2013impact,le2024strack,reps} can cause excessive simultaneous arrivals at a shared link, overflowing buffers and resulting in packet loss throughout the network, as in all-to-all traffic patterns. Coarse splitting fails to balance load effectively, while overly fine splitting overwhelms NICs, breaks hardware offload, and increases CPU overhead, reducing communication-computation overlap in AI training. Designers must therefore navigate the trade-off between effective load balancing and hardware efficiency.

To address these challenges, we propose RailS, whose primary contribution is resolving the fundamental path-selection conflicts. Our theoretical analysis leverages Theorem~\ref{the1} to show that Rail provides $N$ parallel logical channels between any pair of nodes, quantifying its maximum aggregate bandwidth. Building on this, Theorem~\ref{the2} formulates the minimization of communication time as a min-max load balancing problem and establishes a standard linear programming solution. These results indicate that fully bypassing spine-layer forwarding, which is inefficient for MoE traffic, and exclusively using the $N$ direct-connected rails constitutes the optimal strategy for all-to-all communication. RailS operationalizes this approach by activating these previously underutilized parallel paths, supporting up to 64 Queue Pairs (QPs).

Secondly, RailS addresses the "topological blindness" by exploiting the deterministic Rail topology. Theorem~\ref{the3} shows a high symmetry between sending and receiving loads, so any strategy that balances sending loads also balances receiving loads. This provides a theoretical foundation for distributed scheduling, decomposing the global coordination problem into local subproblems where each node optimizes only its own sending load, thereby reducing communication and synchronization overhead. Consequently, RailS operates fully distributed, with each node executing Longest Processing Time first (LPT) scheduling based on local information while collaboratively achieving a near-optimal global solution.

Finally, RailS addresses the trade-off in splitting granularity through theory-driven traffic partitioning and scheduling. Underlying control coalescing mechanism executes low-level splitting, automatically dividing large flows into fixed-size atomic chunks. LPT approximation algorithm then performs proactive, upper-layer scheduling of them, replacing the default reactive logic. Theorem~\ref{the4} shows that the algorithm’s load imbalance is bounded by the maximum size $w_{\max}$, providing theoretical guidance for selecting an optimal granularity that balances load distribution and hardware efficiency.

We validate RailS in a large-scale, programmable datacenter simulator using a variety of synthetic workloads, including uniform, sparse, and skewed patterns, as well as real-world MoE traces. RailS consistently outperforms baselines: under sparse loads, bus bandwidth improves by 20\%–78\% across different sparsity levels and total completion time decreases by 17\%–78\%; in sender-skewed and receiver-skewed scenarios, it achieves the lowest mean squared error across NICs, indicating optimal load balance. For Mixtral workloads, RailS shortens iteration time by 18\%–40\%. In dense setups, it reduces collective completion time by 8.9\%–44.3\%, and in sparse setups, by 66.1\%–80.4\%. These results demonstrate RailS’s ability to fully exploit architectural parallelism while maintaining consistent load balance across diverse and imbalanced workloads.

\section{Background and Motivation}
\begin{figure*}[t]
    \centering
    \begin{subfigure}{2\columnwidth}
        \centering
        \includegraphics[width=\linewidth]{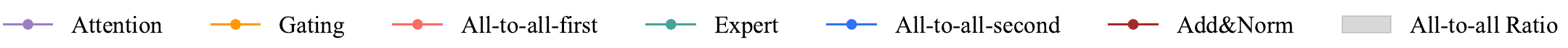}
        \label{moe_legend}
        \vspace{-0.4cm}
    \end{subfigure}
    
    \begin{minipage}[t]{0.247\linewidth} 
        \centering
        \includegraphics[width=\linewidth]{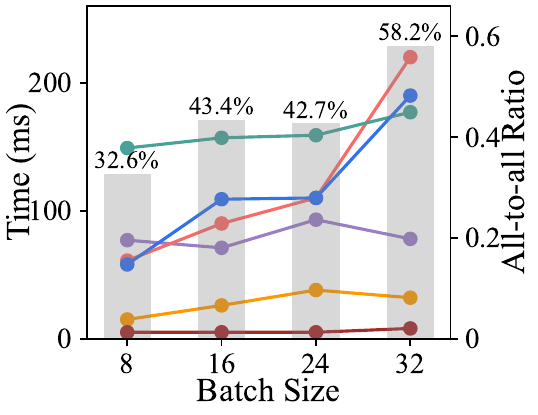}
        \subcaption{Mixtral-MoE} 
        \label{moe_mistral}
    \end{minipage}%
    \begin{minipage}[t]{0.247\linewidth}
        \centering
        \includegraphics[width=\linewidth]{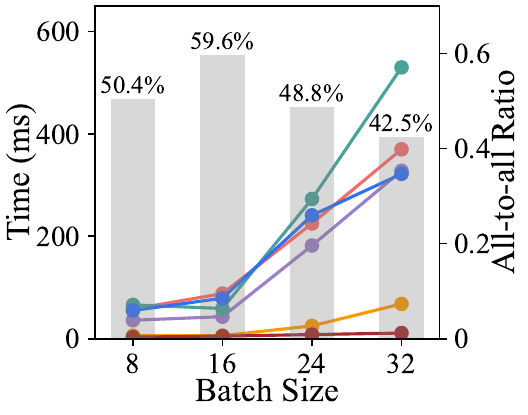}
        \subcaption{LLaMA-MoE}
        \label{moe_llama}
    \end{minipage}%
    \begin{minipage}[t]{0.247\linewidth}
        \centering
        \includegraphics[width=\linewidth]{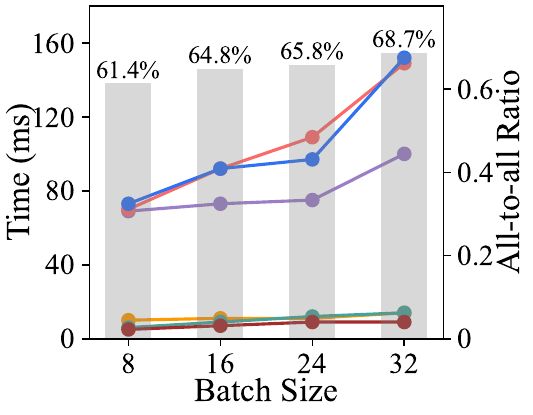}
        \subcaption{Qwen-MoE}
        \label{moe_qwen}
    \end{minipage}%
    \begin{minipage}[t]{0.245\linewidth}
        \centering
        \includegraphics[width=\linewidth]{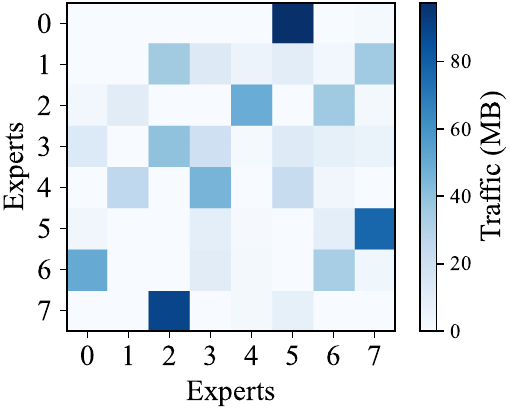}
        \subcaption{Non-uniform all-to-all traffic}
        \label{moe_matrix}
    \end{minipage}
    \caption{Actual measured overhead~\cite{mixnet} of different modules in MoE models under expert parallelism: (a) Mixtral 8×7B, (b) LLaMA 6.7B, and (c) Qwen 14.3B. The all-to-all ratio indicates the measured proportion of iteration time spent on all-to-all communication. (d) Empirically measured traffic matrix of Mixtral 8×7B.} 
    
    \label{moe_traffic}
\end{figure*}

\subsection{Research Background}
\textbf{Model Communication Challenges.} AI training has evolved from deep neural networks (DNNs) with millions of parameters to LLMs with tens of billions, greatly increasing communication demands. Early data-parallel DNNs relied on allreduce for gradient synchronization with limited bandwidth pressure. In contrast, LLMs employ hybrid parallelism including data, tensor, and pipeline modes, where communication volume scales with model size and coordination complexity. In Fig.~\ref{fig:moe}, MoE models further introduce sparsely activated experts, with only a subset active per sample. This produces highly sparse and dynamic all-to-all communication in Fig.~\ref{moe_traffic}, where all-to-all occupies 40\%–60\% of the iteration time, and its load imbalance confirms the impact on performance. Consequently, MoE training reveals the bottlenecks of communication-limited distributed systems.

\textbf{Network Architectures Evolution.} To meet the growing bandwidth and latency demands of AI training, data center networks~\cite{fattree,vl2,dcell,bcube,googleclos,nvidiarail} have evolved from Fat-tree to Spine-leaf and Rail architectures. Three-tier designs suffer from high latency and centralized bottlenecks, while spine–leaf architectures reduce latency over three-tier designs, they fall short under GPU-scale all-to-all workloads, causing congestion and imbalance. The Rail architecture, adopted by NVIDIA, connects multiple NICs to different leaf switches for single-hop GPU paths and parallel bandwidth use. Yet in MoE training, dynamic loads and path constraints still cause NIC congestion and bandwidth underutilization.

\textbf{Multipath Transmission and Load Balancing.} Multipath transmission and load balancing are essential for improving data center communication and supporting distributed node coordination. Early work used flowlet-based TCP splitting, while protocols such as MPTCP~\cite{mptcp} leverage multiple paths at the connection level, though their complexity limits large-scale deployment. With widespread RDMA~\cite{rdma} adoption in AI clusters, communication increasingly relies on single-path low-latency transmission, which restricts subflow coordination and amplifies bursty all-to-all loads in distributed MoE training, reducing global efficiency. Recent frameworks explore multipath optimization to better utilize available network paths, alleviate hotspots, and improve overall throughput and latency in distributed training.

\subsection{Research Motivation}
\textbf{Traditional Topologies and MoE Traffic Incompatibility.} In Spine-leaf networks, ECMP selects cross-leaf paths through static hashing, which performs well for uniform traffic but fails under MoE all-to-all patterns. Sparse expert activations create highly imbalanced traffic that ECMP cannot adapt to, causing NIC and uplink load disparities up to $5–10\times$. This imbalance prolongs critical all-to-all stages and limits MoE training throughput. Existing mitigations, such as disabling NIC congestion control~\cite{gangidi2024rdma,zhao2025deepep} or adopting dual-plane fabrics~\cite{qian2024alibaba}, risk deadlocks~\cite{hu2016deadlocks,mittal2018revisiting} or incur high cost, without resolving the fundamental mismatch between ECMP and sparse traffic.

\textbf{Rail Architecture Communication Bottlenecks.} The Rail architecture connects multiple NICs directly to leaf switches, enabling inherent multipath parallelism. Yet in distributed MoE all-to-all training, current implementations underutilize this potential. RoCE~\cite{roce,rocev2} enforces fixed NIC-level paths, preventing dynamic scheduling across NICs. NCCL~\cite{nccl} supports path selection but only within a single NIC, while UCCL~\cite{uccl_transport} lacks topology-aware scheduling and ignores fixed NIC–leaf mappings. Consequently, cross-NIC bandwidth remains uneven and coordination efficiency constrained, leaving Rail’s architectural advantages underutilized in MoE training.

\textbf{Existing Load Balancing Limitations.} Prior MoE load balancing studies focus on computation or expert placement while neglecting topological constraints and multipath characteristics. Some assume ECMP can evenly distribute traffic but overlook fixed NIC–leaf bindings in Rail, causing subflow mismatches. Others pursue software-based subflow scheduling without considering RDMA offloading, introducing overhead and disrupting communication–computation overlap. Lacking integrated modeling of topology and hardware, these approaches fail to mitigate sparse traffic hotspots, leaving load imbalance unresolved in Rail clusters.

\section{overview}
\label{sec3}
This chapter provides a high-level overview of the proposed traffic scheduling system. First, we describe the core design premises and problem boundaries of the system. Next, we present the overall system design blueprint, including its key mechanisms and optimization objectives. Finally, this chapter summarizes the critical theoretical contributions introduced in subsequent chapters, providing readers with a clear roadmap.

\subsection{Design Premises and Problem Scope}
Before detailing our system design, this section first clarifies the core premises, model assumptions, and problem boundaries of our work.

\begin{figure}[t]
	\centering
	\includegraphics[width=\linewidth]{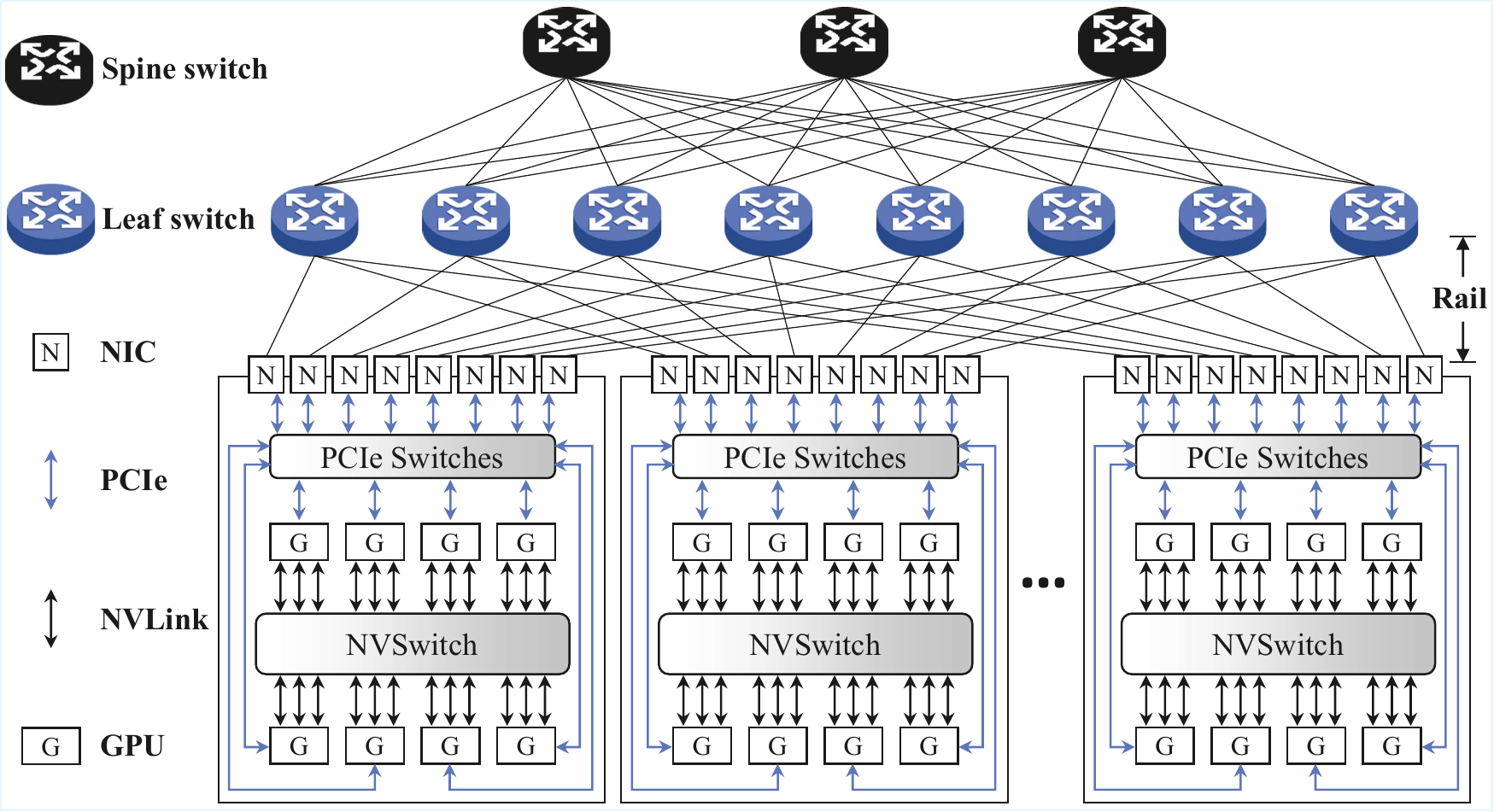}
	\caption{Rail-optimized network topology.}
	\label{fig:topology}
\end{figure}

\textbf{Network Topology.} In Fig.~\ref{fig:topology}, this study focuses on the Rail network architecture, which equips each compute node with multiple NICs, each directly connected to a different leaf switch, providing a hardware foundation for multipath parallel communication. All subsequent theoretical analyses and system designs are built upon the intrinsic structural characteristics of this emerging topology.

\textbf{Traffic Model.} Our optimization target is the key performance bottleneck in current large-scale distributed training: all-to-all communication generated by MoE models. This traffic exhibits significant sparsity and imbalance, rendering traditional static load-balancing mechanisms inefficient. Our theoretical analysis and real-time scheduling are based on the known inter-GPU traffic matrix $D^{(1)}$ and inter-domain traffic matrix $D^{(2)}$.

\textbf{System Abstraction.} At the software level, the system adopts a scalable transport-layer framework that decouples the control and data paths of RDMA NICs. This separation enables flexible, software-defined implementation of transport-layer protocols on the host CPU. The extensibility provides the technical foundation for integrating a custom LPT scheduler to enable proactive load balancing.

\subsection{System Design Overview}
We design and propose RailS, a distributed traffic load balancer specifically optimized for MoE communication under the Rail architecture. The system aims to minimize the communication completion time $T^\star$. Its core idea is to actively shape and schedule sparse, imbalanced flows at each sender node through a set of coordinated mechanisms, thereby approximating globally optimal load balancing.

The RailS's design comprises three core mechanisms forming an efficient splitting-scheduling-spraying pipeline:

\textbf{Flow Splitting.} Serving as the entry point of the pipeline, this mechanism decomposes large, coarse-grained messages generated by the upper-layer applications into a series of smaller, atomic flow units. This preprocessing step is crucial for fine-grained load balancing and for reducing the maximum flow size $w_{\max}$.

\textbf{LPT Scheduling.} Acting as the brain of the system and the core of the proactive spraying, this mechanism differs from traditional strategies that rely on passive feedback such as network latency. Using an LPT-based scheduler, it deterministically assigns each flow to the optimal target track based on the sizes of all pending atomic flows and the real-time local load state of each track, thereby actively constructing a highly balanced flow allocation.

\textbf{Multipath Spraying.} This mechanism provides the macroscopic strategy for flow distribution. It distributes the split atomic flows across all $N$ available parallel communication tracks at each node. Unlike static-hash-based ECMP, spraying mechanism is fully dynamic and policy-driven.

Through the coordinated operation of these mechanisms, RailS effectively decomposes a complex global optimization problem into distributed local decisions that can be executed independently and efficiently at each node, ultimately translating the theoretical insights from Chapter 4 into practical system performance.

\subsection{Core Theoretical Contributions Roadmap}
\begin{figure}[t]
	\centering
	\includegraphics[width=\linewidth]{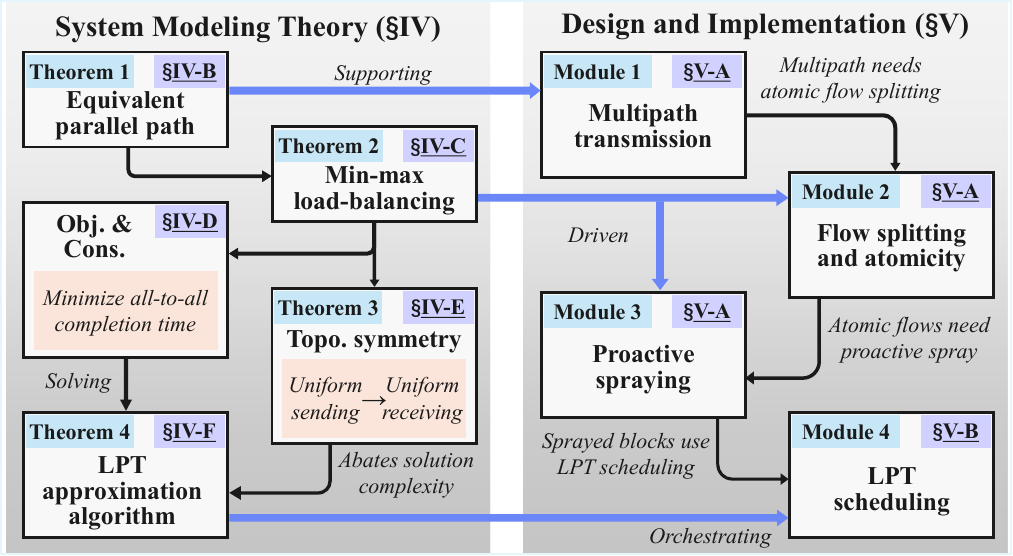}
	\caption{Overview of system workflow.}
	\label{fig:overview}
\end{figure}

This section outlines how a series of interconnected theoretical analyses establish the foundation for the subsequent system design of RailS. Fig.~\ref{fig:overview} illustrates the workflow.

\textbf{Question 1 ($\S$\ref{sec42}): What is the Theoretical Communication Capacity of the Rail Architecture?}
Our analysis begins by rigorously defining the fundamental capability of the Rail network topology. Using the max-flow min-cut method, we prove that the architecture provides $N$ parallel, equal-capacity logical channels for arbitrary inter-domain communication (Theorem~\ref{the1}). This result not only quantifies the system’s bandwidth upper bound but also establishes the feasibility and necessity of multipath load balancing.

\textbf{Question 2 ($\S$\ref{sec43}): How to Formulate the Time Minimization Problem?}
Having established the network capacity, we reformulate the core optimization objective of minimizing all-to-all communication time into a mathematical framework. We prove that this complex dynamic time optimization problem is equivalent to a more tractable min-max load-balancing problem (Theorem~\ref{the2}). This equivalence provides a clear and actionable optimization objective for designing allocation strategies.

\textbf{Question 3 ($\S$\ref{sec45}): What is the Optimal State for Load Balancing?}
To define an “ideal target” for practical scheduling algorithms, we first solve the problem under the idealized assumption that flows are arbitrarily divisible. We prove that the optimal strategy admits a concise closed-form solution, namely a perfectly uniform distribution across $N$ rails (Theorem~\ref{the3}). This result not only reveals the intrinsic symmetry between sending and receiving loads in the Rail architecture but also sets the theoretical performance ceiling for all practical approximation algorithms.

\textbf{Question 4 ($\S$\ref{sec46}):  How to Approximate the Optimum with Discrete Flows?}
Finally, we address the more realistic scenario of indivisible (atomic) flows. We propose the LPT approximation algorithm and prove that it efficiently approaches the ideal uniform state established in Step 3, with a guaranteed performance bound where the mean squared error is upper bounded by $w_{\max}^2$ (Theorem~\ref{the4}). This provides a solid theoretical basis for implementing efficient, low-overhead local scheduling in distributed nodes.

\section{Problem Formulation}
\label{sec4}
In this chapter, we model the Rail-based network topology and reformulate the all-to-all communication optimization as an equivalent load-balancing problem. We derive closed-form optimal strategies for both continuous and atomic flow scenarios and propose an efficient approximation algorithm with proven performance guarantees, establishing a solid theoretical foundation for system design.

\subsection{System Model}
\label{sec41}

We consider a large-scale GPU cluster system composed of $M$ computing domains. The system is communication-constrained and supports distributed training of LLMs and MoE models. Each domain contains $N$ GPUs interconnected via intra-domain networks. Any GPU in the system can be uniquely identified by a tuple $(d, n)$, where $d \in \{1, \dots, M\}$ is the domain index and $n \in \{1, \dots, N\}$ is the GPU index within that domain. Each GPU $(d, n)$ is connected to a dedicated network interface card, denoted as $\mathrm{NIC}_{d,n}$.

The inter-domain communication architecture follows a Rail-based design. Specifically, for any fixed intra-domain index $n$, the $n$-th NIC of all domains (i.e., $\mathrm{NIC}_{1,n}, \mathrm{NIC}_{2,n}, \dots, \mathrm{NIC}_{M,n}$) connects to the same Leaf switch $S_n$. Hence, the system contains $N$ such switches, forming $N$ parallel communication ``rails.'' The Spine and Leaf layers remain fully connected.

To precisely analyze this system, we define the following key concepts.

\textbf{Network Link Rates.}
Two types of link rates are considered. The intra-domain forwarding rate $R_1$ represents the data exchange rate among GPUs within a single domain. The inter-domain forwarding rate $R_2$ corresponds to the forwarding rate of domain NICs or switches and reflects the data processing capability of switch ports.

\textbf{Traffic Matrix.} To quantify communication demands, we define two levels of traffic matrices. The GPU-to-GPU traffic matrix $D^{(1)}$ has elements $D^{(1)}_{(d,n),(f,m)}$ representing the traffic from source GPU $(d,n)$ to destination GPU $(f,m)$. Aggregating these yields the domain-to-domain traffic matrix $D^{(2)}$, where each element $D^{(2)}_{d,f}$ denotes the total traffic from domain $d$ to domain $f$:
\begin{equation}
D_{d,f}^{(2)} = \sum_{n=1}^{N} \sum_{m=1}^{N} D_{(d,n),(f,m)}^{(1)}
\end{equation}

\textbf{Allocation Matrix.} The allocation matrix $P$ is a three-dimensional decision variable describing the routing of traffic. Its elements $P_{k,f,n}$ denote the proportion of total traffic from source domain $k$ to destination domain $f$ assigned to the $n$-th communication rail. The matrix and constraints are given by:
\begin{equation}
P = [P_{k,f,n}] \in \mathbb{R}^{M \times M \times N}
\end{equation}
\begin{equation}
\text{s.t.} \quad 
\sum_{n=1}^{N} P_{k,f,n} = 1, \quad P_{k,f,n} \ge 0, \quad \forall k,f,n
\end{equation}

\textbf{Load Matrix.} Given an allocation matrix $P$, the communication load on each NIC can be computed. We define the sending load matrix $\mathbf{S} \in \mathbb{R}^{M \times N}$ and receiving load matrix $\mathbf{R} \in \mathbb{R}^{M \times N}$:
\begin{equation}
\mathbf{S}_{k,n} = \sum_{f=1}^{M} D^{(2)}_{k,f} P_{k,f,n}
\end{equation}
\begin{equation}
\mathbf{R}_{f,n} = \sum_{k=1}^{M} D^{(2)}_{k,f} P_{k,f,n}
\end{equation}

\textbf{All-to-All Completion Time.} In the all-to-all communication pattern, the total completion time $T$ is defined as the time from the start of communication until all inter-domain data transfers are finished. It is determined by the most heavily loaded NIC (either sending or receiving) and serves as the key metric of system communication performance. Minimizing the time $T^\star$ is one of the main optimization objectives.

\textbf{Load Balancing Mean Squared Error.} To quantify the deviation between actual NIC loads and the ideal target within a domain, we define the mean squared error (MSE). Let the ideal target load $T_{\text{opt}}$ be the average load of the domain, and let $L_j$ denote the load assigned to the $j$-th NIC. The MSE is:
\begin{equation}
\text{MSE} = \frac{1}{N} \sum_{j=1}^{N} (L_j - T_{\text{opt}})^2
\end{equation}

\subsection{Inter-Domain Communication Capacity Analysis}
\label{sec42}

We now analyze the fundamental limits of inter-domain communication in the Rail-based network. This section quantifies the theoretical upper bound of cross-domain throughput and establishes its equivalence-in-use to a set of parallel logical rails, which provides the analytical foundation for subsequent flow allocation and optimization.

\begin{theorem}
\label{the1}
Under the defined system model and assuming that the intra-domain forwarding rate is higher than the inter-domain forwarding rate ($R_1 > R_2$), the maximum achievable unidirectional throughput between any two distinct domains $k$ and $f$ is:
\begin{equation}
Cap_{k \to f} = N \cdot R_2
\end{equation}
\end{theorem}

\begin{proof}
We represent the Rail-based interconnect as a directed capacitated graph $G=(V,E)$. Each node corresponds to either a GPU, a NIC, or a switch, and each directed edge $e\in E$ has transmission capacity $c_e$. Let $\mathbf{c}=(c_e)_{e\in E}$ and $\mathbf{f}=(f_e)_{e\in E}$ satisfy:
\begin{equation}
0 \le f_e \le c_e, \quad \forall e\in E
\label{eq:cap-const}
\end{equation}

For inter-domain analysis, each domain $d$ is modeled as a cluster of $N$ GPUs with dedicated NICs $(\mathrm{NIC}_{d,1}, \dots, \mathrm{NIC}_{d,N})$. We contract the source domain $k$ to a super-source $s$ and the destination domain $f$ to a super-sink $t$; the validity of this contraction will be justified by a no-saturation argument under $R_1>R_2$ below. Let $B\in\mathbb{R}^{|V|\times|E|}$ be the node–edge incidence matrix. Flow conservation is:
\begin{equation}
B\mathbf{f}=\mathbf{b},\qquad
\mathbf{b}=\bigl( \ldots, \underbrace{+x}_{v=s}, \ldots, \underbrace{-x}_{v=t}, \ldots \bigr)^\top,
\label{eq:flow-cons}
\end{equation}
where $x$ denotes the unidirectional throughput from $k$ to $f$. The feasible region is:
\begin{equation}
\mathcal{F}=\left\{(x,\mathbf{f})\;:\;
B\mathbf{f}=\mathbf{b},\ 
\mathbf{0}\le \mathbf{f}\le \mathbf{c}
\right\}
\label{eq:feasible-region}
\end{equation}

\paragraph{Cut-Based Upper Bound}
For any $s$–$t$ cut $(C,\bar C)$ with $s\in C$, $t\in\bar C$, the forward cut-set is $\delta^+(C)$. By the Max-Flow–Min-Cut theorem, it has:
\begin{equation}
x \le \mathrm{Cap}(C):=\sum_{e\in \delta^+(C)} c_e
\label{eq:generic-cut}
\end{equation}
Choose $C$ to include all nodes of domain $k$. The forward edges leaving $C$ are exactly the $N$ inter-domain uplinks:
\begin{equation}
E_{k\to \mathrm{fabric}}=\bigl\{e_n:\ \mathrm{NIC}_{k,n}\to S_n \bigr\}_{n=1}^N
\end{equation}
where $S_n$ denotes the leaf switch attached to $\mathrm{NIC}_{k,n}$ (and symmetrically to $\mathrm{NIC}_{f,n}$). Each $e_n$ has capacity $R_2$, so
\begin{equation}
\mathrm{Cap}(C)=\sum_{n=1}^{N} c_{e_n} = N\!\cdot\! R_2
\label{eq:cut-value}
\end{equation}
Therefore: 
\begin{equation}
Cap_{k\to f}\le N\!\cdot\! R_2
\label{eq:upper-NR2}
\end{equation}

\paragraph{Lower Bound}
Let $\mathcal{P}$ denote the set of simple $s\to t$ paths. We explicitly construct $N$ rail-aligned paths:
\begin{equation}
\Pi_n:
s\to \mathrm{NIC}_{k,n}\to S_n\to \mathrm{NIC}_{f,n}\to t,
\:\: n=1,\ldots,N
\label{eq:paths}
\end{equation}
where each path $\Pi_n$ uses a distinct NIC pair $(\mathrm{NIC}_{k,n}, \mathrm{NIC}_{f,n})$ and its associated leaf $S_n$. The NIC–leaf attachments are fixed per rail, so intra-domain edges used by different rails are edge-disjoint.

Assign rate $\alpha_n = R_2$ to each $\Pi_n$. Define the path–edge incidence variable $a_{e,n}\in\{0,1\}$, which equals $1$ if edge $e$ lies on path $\Pi_n$ and $0$ otherwise. The induced edge load is:
\begin{equation}
f_e = \sum_{n=1}^{N} \alpha_n a_{e,n}
     = \sum_{n=1}^{N} R_2\, a_{e,n}
\label{eq:edgeflow}
\end{equation}
By construction, all inter-domain edges are edge-disjoint across $\{\Pi_n\}_{n=1}^N$, and each such edge has capacity $R_2$, hence $f_{e_n}=R_2\le c_{e_n}=R_2$.

Let $E_{\mathrm{intra}}$ denote the set of intra-domain edges. For intra-domain segments on $\Pi_n$, every traversed edge has capacity at least $R_1$ and carries only the rate of its own rail $\alpha_n=R_2$. Therefore, for each $n=1,\ldots,N$ and each intra-domain edge $e \in \Pi_n \cap E_{\mathrm{intra}}$, we have:
\begin{equation}
f_e = R_2,\ \ c_e \ge R_1 > R_2 \ \Rightarrow\ f_e < c_e
\label{eq:intra-nosat}
\end{equation}

No intra-domain edge saturates under $R_1>R_2$. Hence, contracting each domain to a super-node preserves the max-flow value. The aggregate throughput achieved by $\{\Pi_n\}$ is:
\begin{equation}
x=\sum_{n=1}^{N} \alpha_n = N\!\cdot\! R_2
\label{eq:achieved}
\end{equation}
which proves:
\begin{equation}
Cap_{k\to f}\ge N\!\cdot\! R_2
\label{eq:lower-NR2}
\end{equation}
Combining \eqref{eq:upper-NR2} and \eqref{eq:lower-NR2} yields the equality.

\paragraph{Engineering Note}
The theorem only assumes $R_1>R_2$, whereas practical systems usually have $R_1\!\gg\!R_2$ because intra-domain networks (e.g., NVLink) exceed NIC-based inter-domain links by a large margin, often close to an order of magnitude. Hence, inter-domain links form the bottleneck in practice, validating our parallel-rails view.
\end{proof}

\textbf{Summary.} Rail-based network provides $N$ independent logical rails, each with capacity $R_2$, for any domain pair $(k,f)$. Therefore, the overall system capacity scales linearly with $N$, and subsequent optimization focuses on how to allocate traffic across these $N$ rails to achieve global load balance and minimize all-to-all completion time.

\subsection{Completion Time as a Load Balancing Problem}
\label{sec43}
Building on Theorem~\ref{the1}, which establishes that inter-domain communication relies on $N$ parallel rails, we now analyze the global all-to-all communication pattern. The objective is to minimize the overall completion time $T^\star$. Since direct optimization of $T^\star$ is intractable, the problem is reformulated as an equivalent resource allocation problem in the form of load balancing across rails, which provides a more tractable analytical framework.

The all-to-all communication demand is described by the inter-domain traffic matrix $D^{(2)}$, while a specific flow allocation is characterized by the allocation matrix $P$. These jointly determine the per-domain sending and receiving loads $\mathbf{S}_{d,n}$ and $\mathbf{R}_{d,n}$ on each rail $n$. The system completion time is ultimately governed by the maximum processing time among all such loads. The following theorem formalizes this relationship between completion time and load distribution.

\begin{theorem}
\label{the2}
Under the defined model and given the traffic matrix $D^{(2)}$, the minimum completion time $T^\star$ of the all-to-all communication is equivalent to solving a min-max load balancing problem over all NICs in both sending and receiving directions. If the rate of all inter-domain links is $R_2$, then:
\begin{equation}
T^{\star} = \frac{1}{R_2} \min_{P} \left\{ \max_{k,f,n} \left( \mathbf{S}_{k,n}, \mathbf{R}_{f,n} \right) \right\}
\end{equation}

where $\min_{P}$ denotes the optimization over the strategy space formed by all feasible allocation matrices $P$, and $\max_{k,f,n}$ indicates the maximum taken over all sending and receiving loads.
\end{theorem}

\begin{figure*}[th]
    \centering
    \includegraphics[width=1\textwidth]{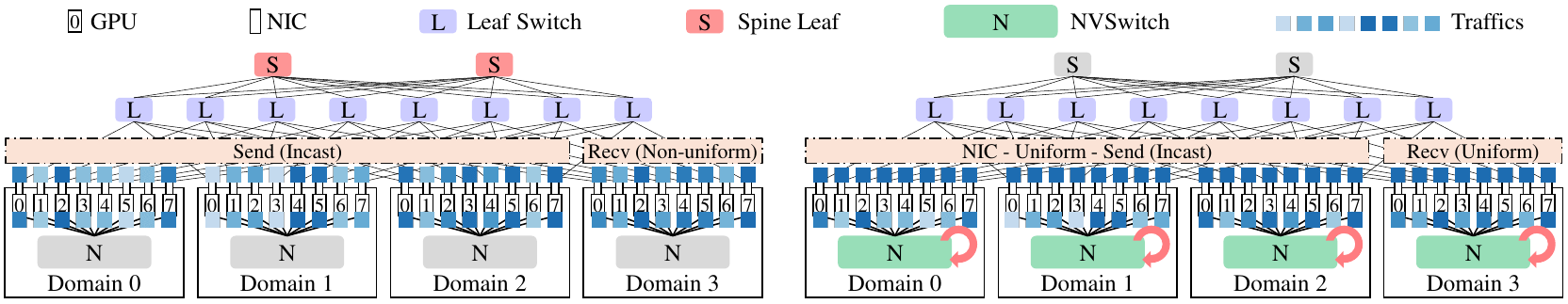}
    \caption{Illustration of symmetry mechanism. The left figure does not leverage the topology. The right figure leverages the intra-domain forwarding and topological properties, where uniform NIC sending results in uniform receiving.}
    \label{fig:symmetry}
\end{figure*}

\begin{proof} 
This proof aims to demonstrate that any achievable completion time $T$ is necessarily bounded below by the maximum load across all NICs, and that this lower bound can be attained by an optimal allocation.

For any given allocation strategy $P$, the corresponding load matrices $\mathbf{S}$ and $\mathbf{R}$ are fully determined. Completion of the all-to-all communication task implies that all sending and receiving operations on every track for every domain must be finished. Therefore, the total completion time $T$ must be at least as large as the time required for any individual NIC to complete its assigned workload.

For any $\mathrm{NIC}(d,n)$, the time to complete its sending load $\mathbf{S}_{d,n}$ is $\mathbf{S}_{d,n}/R_2$, and the time to complete its receiving load $\mathbf{R}_{d,n}$ is $\mathbf{R}_{d,n}/R_2$. Hence, we have:
\begin{equation}
T \geq \frac{\mathbf{S}_{d,n}}{R_2} \quad \text{and} \quad T \geq \frac{\mathbf{R}_{d,n}}{R_2}
\end{equation}

Considering all NICs, the completion time under this strategy must satisfy:
\begin{equation}
T \geq \frac{1}{R_2} \max_{k,f,n} \left( \mathbf{S}_{k,n}, \mathbf{R}_{f,n} \right)
\end{equation}

This inequality holds for all feasible allocation strategies $P$. Therefore, the global minimum completion time $T^\star$ cannot be smaller than the optimal value obtained by minimizing this maximum load over all strategies:
\begin{equation}
T^{\star} \geq \frac{1}{R_2} \min_{P} \left\{ \max_{k,f,n} \left( \mathbf{S}_{k,n}, \mathbf{R}_{f,n} \right) \right\}
\end{equation}

The above derivation establishes a lower bound. When traffic can be continuously divided, the search for an optimal allocation matrix $P$ minimizing the load $\max(\mathbf{S}, \mathbf{R})$ reduces to a linear programming problem. Standard linear programming theory ensures the existence of an optimal allocation $P^\star$ whose induced maximum load attains the min–max optimum. Under this allocation, the system completion time reaches the lower bound, and the inequality becomes an equality.

In summary, the problem of minimizing the all-to-all time is rigorously proven to be equivalent to minimizing the maximum sending and receiving load across all NICs.
\end{proof} 

\textbf{Summary.} This theorem transforms the dynamic performance metric into a optimizable load objective, guiding the search for an optimal traffic allocation strategy $P$ that minimizes the maximum NIC load. To solve this min-max problem systematically, it must first be formulated as a mathematical program. The next section presents the linear programming formulation that underlies the optimal allocation strategy.

\subsection{Objectives and Constraints}
\label{sec44}
Based on the conclusions of the previous section, we formulate the min-max problem of all-to-all minimum completion time as a standard linear programming (LP) model to facilitate formal solution. The model takes the allocation matrix $P$ as the core decision variable and introduces an auxiliary variable $t$ representing the upper bound of all loads.

The optimization problem is formulated as:
\begin{equation}
\begin{aligned}
\min_{P,t} \quad & t \\
\text{s.t.} \quad & \sum_{f=1}^{M} D_{k,f}^{(2)} \cdot P_{k,f,n} \leq t, \quad \forall k, n \\
& \sum_{k=1}^{M} D_{k,f}^{(2)} \cdot P_{k,f,n} \leq t, \quad \forall f, n \\
& \sum_{n=1}^{N} P_{k,f,n} = 1, \quad \forall k, f \\
& P_{k,f,n} \geq 0, \quad \forall k, f, n
\end{aligned}
\end{equation}

In this LP formulation, the auxiliary variable $t$ represents the objective to be minimized and is defined as the global upper bound of all NIC sending loads $\mathbf{S}_{k,n}$ and receiving loads $\mathbf{R}_{f,n}$. Minimizing $t$ therefore corresponds to minimizing the system’s bottleneck load. The remaining constraints guarantee a valid allocation matrix $P$, ensuring complete traffic assignment (normalization) and correct directionality (non-negativity). As both the objective and constraints are linear, the problem can be solved optimally using standard LP methods.

This LP formulation provides a numerical solution to the problem, while the next section will further explore and present a more insightful analytical solution.

\subsection{Topological Symmetry of Load}
\label{sec45}
The all-to-all minimum completion time problem has been formulated as a linear program. Here we derive a closed-form analytical solution and prove its optimality. The key insight is that in the Rail-based architecture, uniform sending loads inherently yield uniform receiving loads in Fig.~\ref{fig:symmetry}.

\begin{theorem}
\label{the3}
In the studied system, any allocation strategy that achieves uniformization of sending loads is globally optimal. Specifically, if for every source domain $k$, the sending load is uniformly distributed as:
\begin{equation}
\mathbf{S}_{k,n} = \frac{1}{N} \sum_{f=1}^{M} D_{k,f}^{(2)}, \quad \forall n = 1, \ldots, N
\end{equation}
then the receiving load for any destination domain $f$ is automatically uniformized:
\begin{equation}
\mathbf{R}_{f,n} = \frac{1}{N} \sum_{k=1}^{M} D_{k,f}^{(2)}, \quad \forall n = 1, \ldots, N
\end{equation}
This strategy simultaneously minimizes both sending and receiving bottlenecks, thereby achieving the all-to-all minimum completion time $T^\star$.
\end{theorem}

\begin{proof}
The core of this proof relies on the system structure revealed in Theorem~\ref{the1}. Communication from any source domain $k$ to destination domain $f$ is equivalent to $N$ parallel, independent paths, with endpoints $\mathrm{NIC}_{k,n}$ and $\mathrm{NIC}_{f,n}$ for $n = 1, \dots, N$. This implies that traffic sent from $\mathrm{NIC}_{k,n}$ along the $n$-th track is uniquely directed to $\mathrm{NIC}_{f,n}$.

We first construct a sending-uniform strategy. This strategy evenly distributes the total outgoing traffic of each source domain $k$, $\sum_{f=1}^{M} D^{(2)}_{k,f}$, across its $N$ NICs. Specifically, the sending load on the $n$-th NIC is set equal to the average load. By adopting the allocation strategy:
\begin{equation}
P^\star_{k,f,n} = \frac{1}{N}
\end{equation}
we obtain the following equation:
\begin{equation}
\mathbf{S}_{k,n} = \sum_{f=1}^{M} D_{k,f}^{(2)} P_{k,f,n}^{\star} = \frac{1}{N} \sum_{f=1}^{M} D_{k,f}^{(2)}
\end{equation}

This strategy successfully uniformizes the sending load, thereby minimizing the maximum processing time among all sending tasks.

Next, we examine the impact of this strategy on receiving loads. The receiving load $\mathbf{R}_{f,n}$ on the $n$-th track for any destination domain $f$ is the sum of the traffic from all source domains $k$ through that track. Based on the one-to-one correspondence of paths in the Rail architecture, it has:
\begin{equation}
\mathbf{R}_{f,n} = \sum_{k=1}^{M} D_{k,f}^{(2)} P_{k,f,n}^{\star} = \frac{1}{N} \sum_{k=1}^{M} D_{k,f}^{(2)}
\end{equation}
This shows that the receiving load is automatically uniformized, with each receiving NIC carrying exactly the average of the domain's total incoming traffic.

Thus, a simple sending-uniform strategy, due to the intrinsic system architecture, inevitably leads to uniform receiving loads. Uniform sending minimizes the sending bottleneck, while the resulting uniform receiving minimizes the receiving bottleneck. Since this single strategy simultaneously optimizes both sending and receiving loads, it is globally optimal and achieves the all-to-all minimum completion time $T^\star$.

Consequently, the optimal allocation matrix achieving this state has a simple closed-form solution:
\begin{equation}
P_{k,f,n}^{\star} = \frac{1}{N}, \quad \forall k, f, n
\end{equation}
which satisfies the normalization constraint:
\begin{equation}
\sum_{n=1}^{N} P_{k,f,n}^{\star} = 1
\end{equation}

The proof is complete.
\end{proof}

\textbf{Summary.} This theorem provides the optimal allocation strategy under the ideal assumption that traffic can be arbitrarily split. However, in practical systems, traffic often exists as discrete units that cannot be fully subdivided. The next section investigates how to effectively approximate this ideal uniform allocation strategy under such discrete constraints.

\subsection{LPT-based Load Balancing for Atomic Flows}
\label{sec46}
The previous section derived the optimal allocation under the ideal assumption of arbitrarily divisible traffic. In practice, however, communication between GPUs is scheduled as indivisible units, making continuous allocation infeasible. Even when flows are split, they remain discrete, as large flows are only divided into several smaller ones (packets or flowlets) rather than continuous segments (bytes or bits).

This section addresses the load balancing problem under such discrete constraints. We aim to minimize the mean squared error between the actual loads and the ideal uniform loads guided by Theorem~\ref{the3}. To this end, we propose an efficient approximate allocation algorithm and provide theoretical guarantees for its performance.

We assume that each flow is indivisible or has been split into the smallest atomic units. The inter-domain total traffic matrix $D^{(2)}$ is known, i.e., the flow between each source domain $k$ and destination domain $f$, $D^{(2)}_{k,f}$, is given. Based on a fixed allocation matrix $P$, let $p_n$ denote the fraction of traffic ideally assigned to the $n$-th NIC. Then the target load for each flow on NIC $n$ is defined as:
\begin{equation}
T_{k,n} = \sum_{f=1}^{M} D^{(2)}_{k,f} p_n
\end{equation}

Each GPU-to-GPU flow must be assigned entirely to a single NIC. The problem can thus be formalized as allocating a set of discrete flows with weights $w_i$ to $N$ NICs, such that the actual load on each NIC
\begin{equation}
L_j = \sum_i w_i x_{i,j}, \quad x_{i,j} \in \{0,1\}
\end{equation}
is as close as possible to its target load $T_j$. The binary assignment variables $x_{i,j}$ satisfy:
\begin{equation}
x_{i,j} \!=\!
\begin{cases}
1, \text{flow } i \text{ on NIC } j\\
0, \text{otherwise}
\end{cases}
,\sum_{j=1}^{N} x_{i,j} = 1, \ \forall i
\end{equation}

To quantify the load deviation, we introduce the MSE as the metric:
\begin{equation}
\text{MSE} = \frac{1}{N} \sum_{j=1}^{N} (L_j - T_j)^2
\end{equation}

Under this definition, the GPU-to-GPU flow allocation problem with fixed allocation fractions $p_n$ can be formalized as a combinatorial optimization problem:
\begin{equation}
\begin{aligned}
\min_{x_{i,j}} \quad & \frac{1}{N} \sum_{j=1}^{N} (L_j - T_j)^2 \\
\text{s.t.} \quad & \sum_{j=1}^{N} x_{i,j} = 1, \quad \forall i, \\
& L_j = \sum_{i} w_i x_{i,j}, \quad \forall j, \\
& x_{i,j} \in \{0,1\}, \quad \forall i,j
\end{aligned}
\end{equation}

This problem is a discrete combinatorial optimization task, aiming to minimize the mean squared deviation between the actual load and the target load on each NIC. Since each flow is assigned to exactly one NIC, directly solving this problem incurs high computational complexity in large-scale systems. To reduce computational overhead, we adopt the LPT for approximate allocation. 

The LPT method first sorts all flows in descending order of weights $w_i$ and then sequentially assigns each flow to the NIC with the currently smallest total load until all flows are allocated. According to Theorem~\ref{the3}, the sending-uniform strategy ensures that the target load $T_j$ already matches the ideal average load $L_\mathrm{opt} = \sum_i w_i / N$, so the LPT algorithm effectively performs discrete allocation while maximizing the utilization of uniformly balanced NIC loads.

\begin{theorem}
\label{the4}
For the discrete-flow load balancing problem described above, when the LPT algorithm is employed for allocation, the MSE between the actual load and the ideal target load is upper-bounded by the square of the largest single flow weight $w_{\max}$:
\begin{equation}
\text{MSE} \le w_{\max}^2
\end{equation}
\end{theorem}

\begin{proof}
Classical LPT theory~\cite{lpt} guarantees that, for the standard discrete load balancing problem, the maximum load $L_\mathrm{max}$ satisfies:
\begin{equation}
L_\mathrm{max} \le L_\mathrm{opt} + w_\mathrm{max}
\end{equation}
where $w_\mathrm{max} = \max_i w_i$ denotes the weight of the largest single flow. The LPT approximation ratio is given by:
\begin{equation}
R = \frac{L_\mathrm{max}}{L_\mathrm{opt}} \le \frac{4}{3} - \frac{1}{3N}
\end{equation}

Consequently, the deviation of the actual load on each NIC from the target load is bounded by:
\begin{equation}
|L_j - L_\mathrm{opt}| \le L_\mathrm{max} - L_\mathrm{opt} \le w_\mathrm{max}, \quad \forall j.
\end{equation}

This directly leads to the following upper bound on MSE:
\begin{equation}
\text{MSE} = \frac{1}{N}\sum_{j=1}^{N} (L_j - L_\mathrm{opt})^2 \le w_\mathrm{max}^2
\end{equation}

Thus, by sorting flows in descending order and greedily assigning them to the NIC with the current minimum load, LPT effectively minimizes the load deviation of each NIC and indirectly controls the MSE. The computational complexity of the algorithm is dominated by sorting and assignment operations, which are $O(F \log F)$ and $O(F N)$, respectively, yielding an overall complexity of $O(F \log F + F N)$, where $F$ denotes the number of flows. Compared to the exponential complexity of an exact ILP solution, LPT provides a significant reduction in computational overhead, making it suitable for large-scale multi-GPU systems.
\end{proof}

\textbf{Summary.} This section extends the continuous-flow theory to practical systems with indivisible flows. Through the LPT algorithm, we show that discrete allocation can still achieve predictable load balancing. The key insight is that load balancing effectiveness depends on the maximum flow granularity, motivating the next focus on designing flow-splitting and scheduling mechanisms that control granularity to approach the theoretical optimum.

\section{Implementation}
\label{sec5}

\begin{figure}[t]
	\centering
	\includegraphics[width=\linewidth]{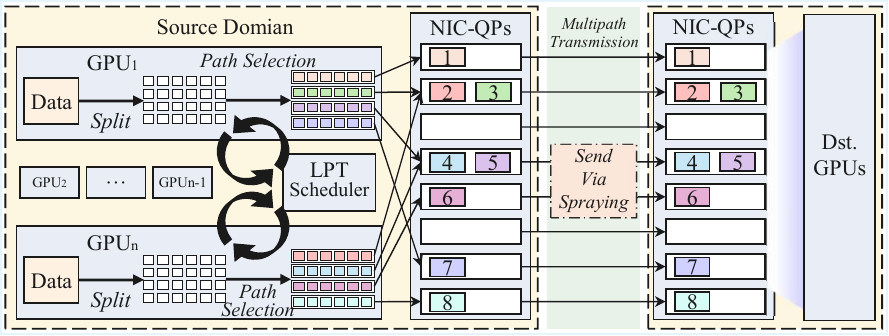}
	\caption{System design and implementation.}  
	\label{fig:design}
\end{figure}

Building upon the theoretical framework established in the previous chapter, this section presents the concrete design and implementation of an all-to-all communication system optimized for the Rail architecture. We first introduce the overall framework adopted to achieve load balancing, encompassing multi-path transmission, flow splitting, and the spraying mechanism. Subsequently, we provide a detailed description of the design and implementation of the LPT scheduler.

\subsection{Multi-Path spraying Framework}
To fully exploit the $N$ parallel communication lanes provided by the Rail architecture, our system employs multi-path packet spraying as the core traffic distribution mechanism in Fig.~\ref{fig:design}. Spraying is an advanced multi-path transmission technique designed to actively disperse the units of a logical data flow (e.g., packets or subflows) across multiple physical paths. For the sparse and uneven all-to-all traffic generated by MoE models, the spraying mechanism provides finer-grained and more dynamic load balancing compared to traditional ECMP static hashing, effectively alleviating link congestion caused by hash collisions on large flows.

\begin{algorithm}[tb]
\small
\caption{MoE distributed training with DP/EP/PP and LPT-based all-to-all communication.}
\label{alg:moe_training}
\KwData{Local GPU $g$; Number of rails $N$; model parameters $\Theta$; Expert assignment $\mathcal{E}$}
\KwResult{Updated model parameters $\Theta$ after one training iteration}

\tcp{Step 0: Preprocessing / Input preparation}
Load mini-batch input data for this GPU.

\tcp{Step 1: Attention and gating computation}
Compute attention outputs: $AttnOut \leftarrow Attention(Inputs, \Theta)$ \\
Compute gating decisions: $Gate \leftarrow Gate(AttnOut)$

\tcp{Step 2: First All-to-All (inputs to experts)}
Select input slices for local experts according to $Gate$ and $\mathcal{E}$ \\
Split inputs into atomic flows $\{w_i\}$ \\
Call \textbf{LPT Scheduler} to assign flows to rails \\
Transmit flows via spraying

\tcp{Step 3: Expert computation (distributed EP)}
Compute local expert outputs $ExpOut_e$ \\
If expert spans multiple GPUs, aggregate partial results

\tcp{Step 4: Second All-to-All (expert outputs aggregation)}
Split expert outputs into atomic flows $\{w_i\}$ \\
Call \textbf{LPT Scheduler} to assign flows to rails \\
Transmit flows via spraying

\tcp{Step 5: Add \& Norm computation}
$Output \leftarrow AddNorm(AttnOut, ExpOut)$

\tcp{Step 6: Output propagation}
$Out_{MoE} \leftarrow Output$ \\
Pass $Out_{MoE}$ to the next pipeline stage

\end{algorithm}

Our system builds on a scalable software transport framework that supports flexible, multi-path data transmission across NICs. Algorithm~\ref{alg:moe_training} presents the procedure for system operation in MoE training. Its primary advantage lies in its successful decoupling of the RDMA NIC’s control and data paths, enabling flexible implementation and innovation of transport-layer protocols in software on the host CPU. This design provides a natural entry point for integrating a custom LPT scheduler. The framework mainly exploits some key features.

\textbf{Multipath Transmission.} The system realizes multipath communication by establishing and managing multiple QPs between each pair of communicating NICs. In our Rail architecture, each node is equipped with $N$ NICs, and this mechanism is used to construct $N$ parallel logical communication lanes, perfectly corresponding to the $N$ parallel channels in our theoretical model. By default, it supports up to 256 QPs for parallelization and load balancing, providing abundant path resources for traffic spraying.

\textbf{Flow Splitting and Atomicity.} Theorem~\ref{the4} indicates that the effectiveness of load balancing is closely related to the granularity of traffic. The underlying control coalescing mechanism naturally supports flow splitting. It segments large application-layer messages into fixed-size data chunks (e.g., 32~KB by default), and makes independent transmission decisions for each chunk. In our system, these generated fixed-size chunks, or small application-layer messages, are treated as indivisible atomic flows. These atomic flows serve as the fundamental units for load balancing operations.

\textbf{Proactive Spraying.} Current spraying strategies are often reactive, such as randomly sampling path RTTs and probabilistically steering traffic to currently lower-latency paths. While these approaches can adapt to network dynamics, they do not necessarily achieve the globally balanced load that the theoretical analysis guarantees as optimal. Our core design principle is to implement a proactive, theory-guided spraying mechanism on top of the multipath framework. Using a deterministic scheduling algorithm, the system directly computes the optimal target path for each atomic flow. The next section presents the key component that realizes the LPT scheduler.

\subsection{LPT Scheduler}

To realize the proactive, load-aware spraying strategy proposed in the previous section, we design and implement the LPT Scheduler for path selection logic. The scheduler is deployed as a lightweight software module on each source domain and serves as the core decision engine for path selection, translating the theoretical algorithm from Chapter~\ref{sec4} into an efficient engineering practice.

\textbf{Architecture and State Management.} The LPT Scheduler runs independently on each sending node. Its core component is a local load state array of size $N$, denoted as $LoadState[N]$. Each entry $LoadState[j]$ maintains the cumulative number of bytes already assigned to the $j$-th communication lane (i.e., the $j$-th local NIC) for the current all-to-all operation. To adapt to the dynamic nature of MoE traffic, the array is reset at the start of each new all-to-all round, ensuring that scheduling decisions are always based on the current communication round's status.

\begin{algorithm}[tb]
    \small
    \caption{LPT scheduler.}\label{alg:lpt_scheduler}
    \KwData{Number of rails $N$; set of local GPUs $\mathcal{G}$}
    \KwResult{Path assignment $\{(w_i, j^*)\}$ for all flows; load MSE}

    \tcp{Initialization before each All-to-All}
    Initialize $LoadState[1 \dots N] \leftarrow 0$. \\
    Initialize empty flow set $\mathcal{W} \leftarrow \emptyset$. \\

    \tcp{Step 1: Flow collection from local GPUs}
    \For{each GPU $g \in \mathcal{G}$}{
        Receive atomic flows $\{w_{g,1}, w_{g,2}, \dots\}$ from GPU $g$. \\
        Tag each $w_{g,k}$ with source ID $g$. \\
        Append them into $\mathcal{W}$. \\
    }

    \tcp{Step 2: LPT sorting}
    Sort $\mathcal{W}$ by descending weight. \\
    Break ties by GPU index. \\

    \tcp{Step 3: Iterative allocation}
    \For{each flow $w_i \in \mathcal{W}$}{
        $j^* \leftarrow \arg\min_{j \in [1,N]} LoadState[j]$. \\
        Record mapping $(w_i, j^*)$ into assignment table. \\
        Notify source GPU of $w_i$ with selected NIC $j^*$. \\
        $LoadState[j^*] \leftarrow LoadState[j^*] + w_i$. \\
    }

    \tcp{Step 4: Spraying execution}
    \For{each assignment $(w_i, j^*)$}{
        Select port $p$ from NIC $j^*$ by round-robin. \\
        Map $(j^*,p)$ to a QP in NIC’s QP set. \\
        Bind $w_i$ to the chosen QP. \\
        Transmit $w_i$ via transport engine. \\
    }

    \tcp{Step 5: Completion \& Feedback}
    \For{each flow $w_i$}{
        Await completion signal from NIC $j^*$. \\
        Confirm delivery to destination GPU. \\
        Update scheduler statistics with actual transmission time and throughput. \\
    }

    \tcp{Step 6: Performance evaluation (MSE)}
    Compute average load:$\mu \leftarrow \frac{1}{N} \sum_{j=1}^N LoadState[j]$\\
    Compute $MSE \leftarrow \frac{1}{N} \sum_{j=1}^N (LoadState[j] - \mu)^2$ \\
    Log $MSE$ for analysis and scheduler tuning. \\
\end{algorithm}

\textbf{Workflow and Implementation.} Upon the initiation of an all-to-all communication by the upper-layer application, the generated traffic is split into a set of atomic flows $\{w_i\}$. The LPT Scheduler then takes over the path selection for these flows, following the LPT algorithm~\ref{alg:lpt_scheduler} strictly:

\begin{itemize}
    \item \textbf{Flow Sorting:} The scheduler first sorts all atomic flows $\{w_i\}$ to be sent from the local node in descending order of their weights.
    
    \item \textbf{Iterative Assignment:} The scheduler then iterates over the sorted flow list. For each flow $w_i$, it performs the following core steps:
    \begin{itemize}
        \item \textit{Path Selection:} Identify the index of the path with the currently minimum cumulative load in the local $LoadState$ array $j^* = \arg\min_{j} LoadState[j]$.
        \item \textit{Decision Output:} Assign $w_i$ to the selected path $j^*$ and return this decision to transport engine.
        \item \textit{State Update:} Immediately update the local load state $LoadState[j^*] \gets LoadState[j^*] + w_i$.
    \end{itemize}
    
    \item \textbf{Spraying Execution:} Upon receiving the path index $j^*$, the transport engine transmits the atomic flow $w_i$ through the set of QPs associated with the $j^*$-th lane. Repeating this process for all atomic flows completes a global, deterministic spraying, systematically distributing the node’s total outbound traffic across the $N$ parallel lanes according to the LPT-optimized allocation.
\end{itemize}

The LPT Scheduler effectively translates the theoretical model of Theorem~\ref{the4} into a fully distributed low-complexity engineering implementation. Unlike naive random spraying, it performs strategic flow assignments aimed at minimizing load variance, thereby fully exploiting the multi-NIC parallelism provided by the Rail architecture, even under the irregular and sparse traffic patterns typical of MoE workloads.

\section{Evaluation}
\label{sec6}

\subsection{Experimental Setup}\label{sec:setup}
\textbf{Simulation Setup.} We implement a scalable datacenter simulation environment on Mininet 2.3.1b4 with Soft-RoCE, using a Rail-optimized topology. The network comprises multiple spine switches and at least eight fully connected leaf switches. Each computing domain connects in parallel to different leaf switches via multiple NICs, forming Rail-based links. The system scales to 128 compute domains with eight GPUs each interconnected through NVLINK. Network forwarding uses Open vSwitch with OpenFlow for fine-grained control, while a Ryu-based distributed controller installs flow rules and manages ECMP routing.

\begin{figure*}[t]
    \centering
    
    \begin{minipage}[t]{0.198\linewidth} 
        \centering
        \includegraphics[width=\linewidth]{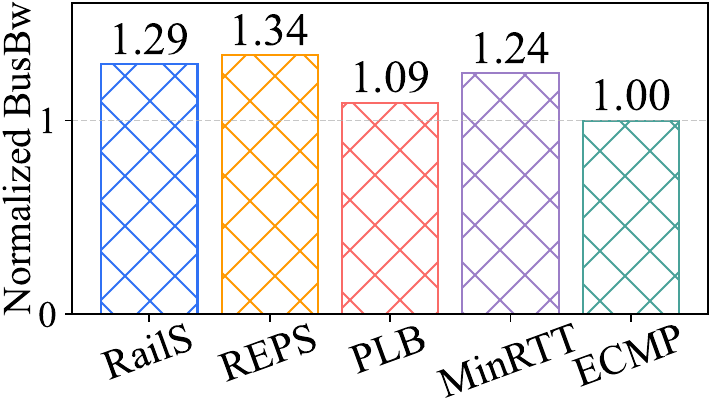}
        \subcaption{Uniform} 
        \label{uniform_busbw}
    \end{minipage}%
    \hfill
    \begin{minipage}[t]{0.198\linewidth}
        \centering
        \includegraphics[width=\linewidth]{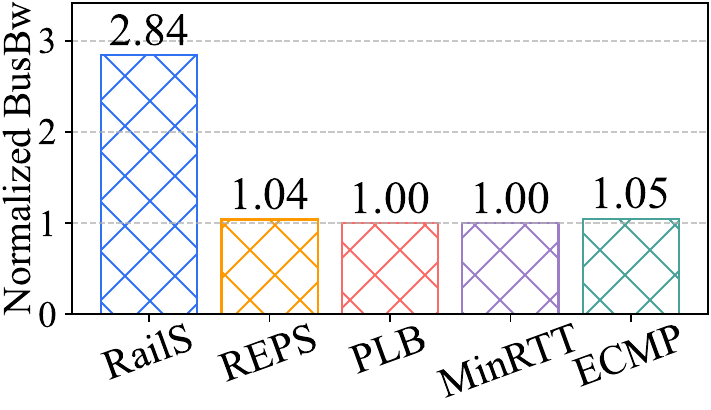}
        \subcaption{Sparse-0.6}
        \label{random_0.4_busbw}
    \end{minipage}%
    \hfill
    \begin{minipage}[t]{0.198\linewidth}
        \centering
        \includegraphics[width=\linewidth]{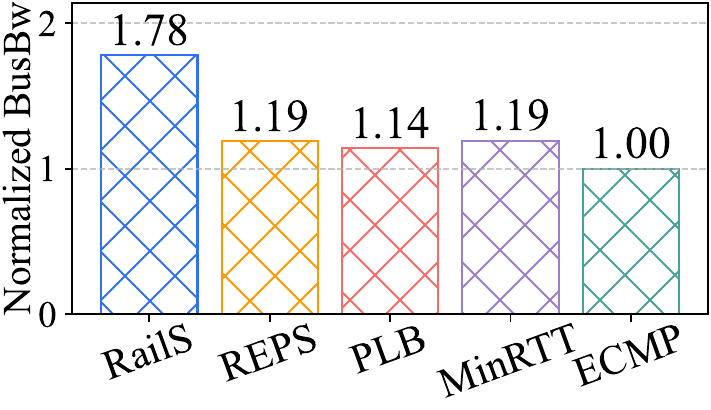}
        \subcaption{Sparse-0.4}
        \label{random_0.6_busbw}
    \end{minipage}%
    \hfill
    \begin{minipage}[t]{0.198\linewidth}
        \centering
        \includegraphics[width=\linewidth]{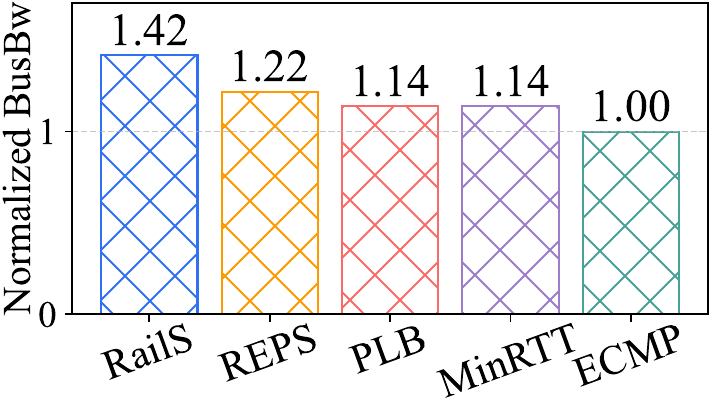}
        \subcaption{Sparse-0.2}
        \label{random_0.8_busbw}
    \end{minipage}
    \hfill
    \begin{minipage}[t]{0.198\linewidth}
        \centering
        \includegraphics[width=\linewidth]{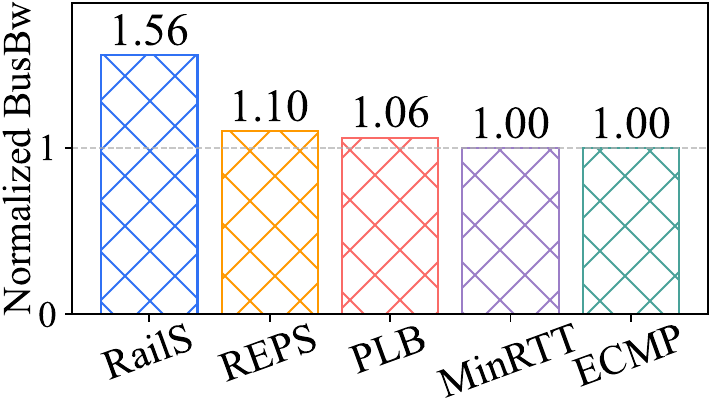}
        \subcaption{Sparse-0 (Fully dense)}
        \label{random_1_busbw}
    \end{minipage}
    \caption{Normalized BusBw of different schemes under various all-to-all workloads: (a) Uniform matrix, (b) Sparse-0.6, (c) Sparse-0.4, (d) Sparse-0.2, and (e) Sparse-0 (fully dense).} 
    
    \label{uniform_sparse_busbw}
\end{figure*}

\begin{figure*}[t]
    \centering
    \begin{subfigure}{2\columnwidth}
        \centering
        \includegraphics[width=\linewidth]{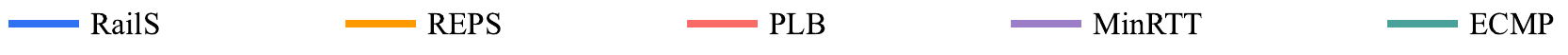}
        \label{cct_legend}
        \vspace{-0.5cm}
    \end{subfigure}
    
    \begin{minipage}[t]{0.198\linewidth} 
        \centering
        \includegraphics[width=\linewidth]{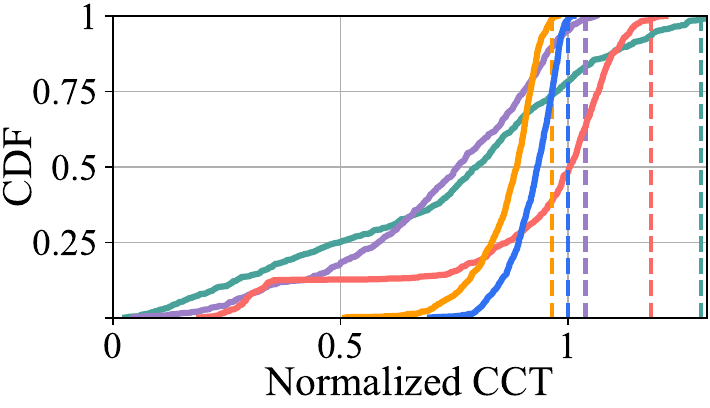}
        \subcaption{Uniform} 
        \label{uniform_fct_cdf}
    \end{minipage}%
    \hfill
    \begin{minipage}[t]{0.198\linewidth}
        \centering
        \includegraphics[width=\linewidth]{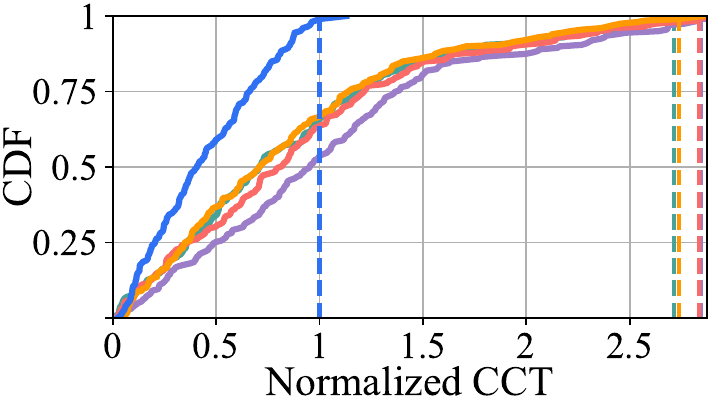}
        \subcaption{Sparse-0.6}
        \label{random_0.4_fct_cdf}
    \end{minipage}%
    \hfill
    \begin{minipage}[t]{0.198\linewidth}
        \centering
        \includegraphics[width=\linewidth]{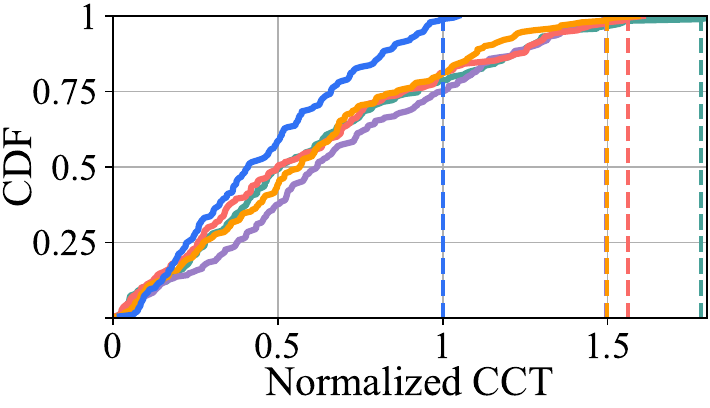}
        \subcaption{Sparse-0.4}
        \label{random_0.6_fct_cdf}
    \end{minipage}%
    \hfill
    \begin{minipage}[t]{0.198\linewidth}
        \centering
        \includegraphics[width=\linewidth]{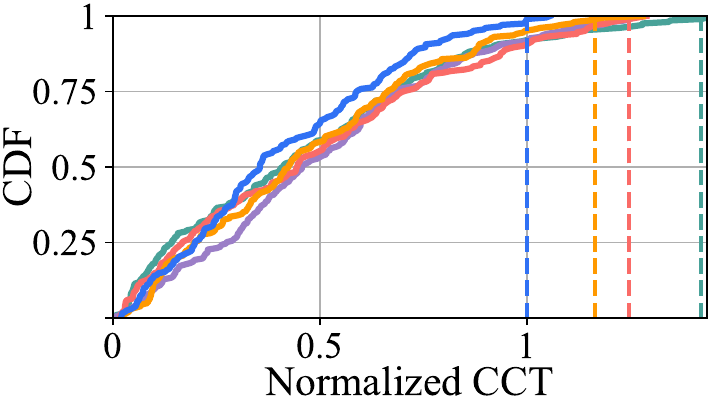}
        \subcaption{Sparse-0.2}
        \label{random_0.8_fct_cdf}
    \end{minipage}
    \hfill
    \begin{minipage}[t]{0.198\linewidth}
        \centering
        \includegraphics[width=\linewidth]{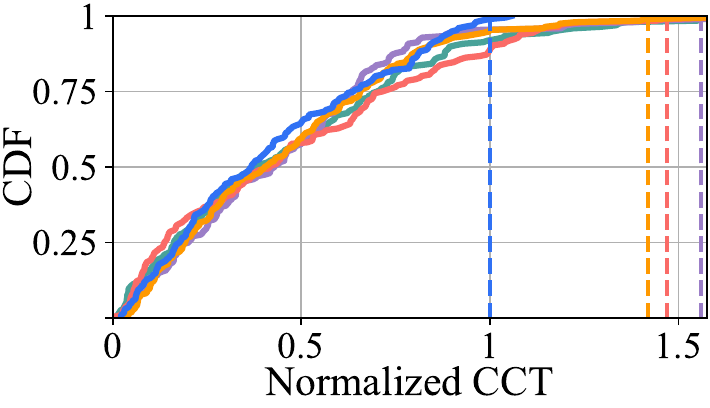}
        \subcaption{Sparse-0 (Fully dense)}
        \label{random_1_fct_cdf}
    \end{minipage}
    \caption{CDF of normalized CCT for different schemes under various all-to-all workloads: (a) Uniform matrix, (b) Sparse-0.6, (c) Sparse-0.4, (d) Sparse-0.2, and (e) Sparse-0 (fully dense).} 
    
    \label{uniform_sparse_cct_cdf}
\end{figure*}

\begin{figure*}[t]
    \centering
    \begin{subfigure}{2\columnwidth}
        \centering
        \includegraphics[width=\linewidth]{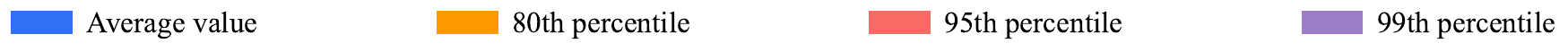}
        \label{legend_percentiles}
        \vspace{-0.4cm}
    \end{subfigure}
    
    \begin{minipage}[t]{0.198\linewidth} 
        \centering
        \includegraphics[width=\linewidth]{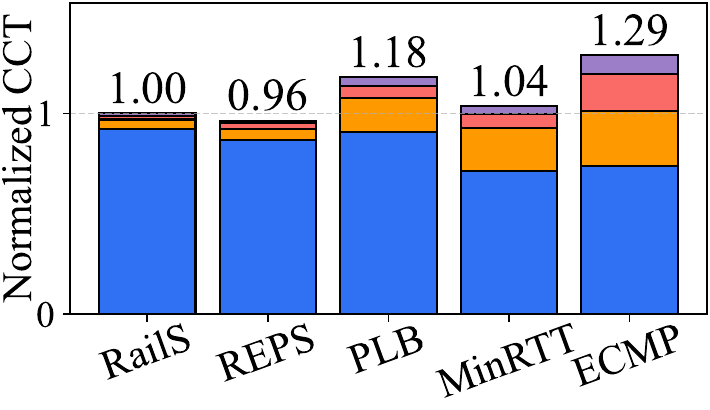}
        \subcaption{Uniform} 
        \label{uniform_fct_perc}
    \end{minipage}%
    \hfill
    \begin{minipage}[t]{0.198\linewidth}
        \centering
        \includegraphics[width=\linewidth]{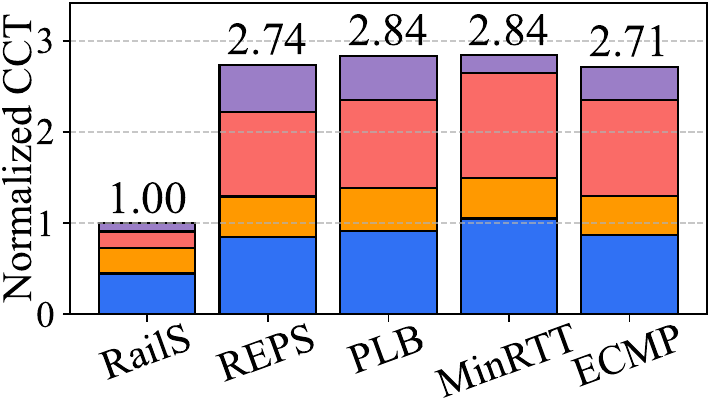}
        \subcaption{Sparse-0.6}
        \label{random_0.4_fct_perc}
    \end{minipage}%
    \hfill
    \begin{minipage}[t]{0.198\linewidth}
        \centering
        \includegraphics[width=\linewidth]{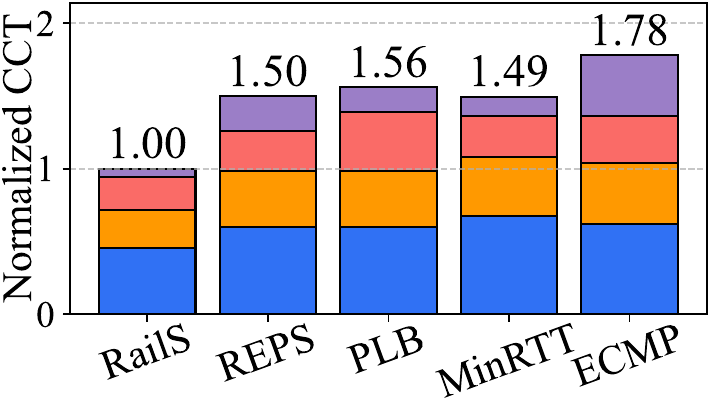}
        \subcaption{Sparse-0.4}
        \label{random_0.6_fct_perc}
    \end{minipage}%
    \hfill
    \begin{minipage}[t]{0.198\linewidth}
        \centering
        \includegraphics[width=\linewidth]{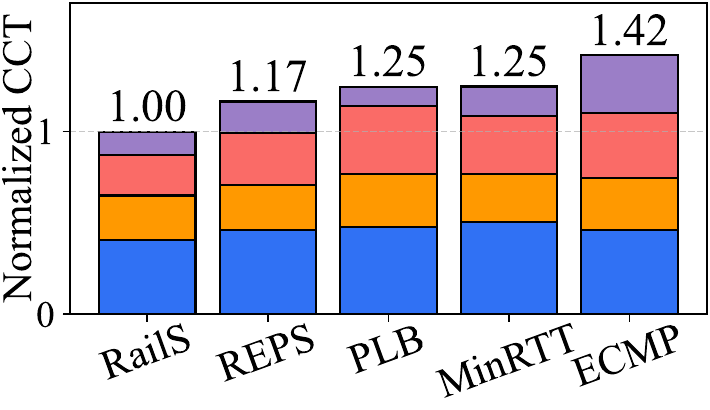}
        \subcaption{Sparse-0.2}
        \label{random_0.8_fct_perc}
    \end{minipage}
    \hfill
    \begin{minipage}[t]{0.198\linewidth}
        \centering
        \includegraphics[width=\linewidth]{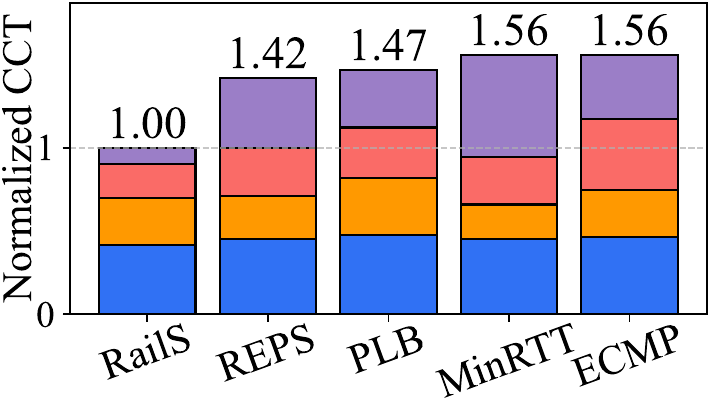}
        \subcaption{Sparse-0 (Fully dense)}
        \label{random_1_fct_perc}
    \end{minipage}
    \caption{Normalized CCT of different schemes under various all-to-all workloads, showing average, 80th, 95th, and 99th percentiles: (a) Uniform matrix, (b) Sparse-0.6, (c) Sparse-0.4, (d) Sparse-0.2, and (e) Sparse-0 (fully dense).} 
    
    \label{uniform_sparse_cct}
\end{figure*}

\textbf{Workloads.} We construct four representative workloads using MoE outputs from the SimAI AICB library~\cite{simai} with controlled token and gating distributions. \textbf{Uniform Workload} sends equal data from each sender to all receivers, yielding balanced link loads without “elephant” or “mice” flows, and serves to measure the gap between achieved and optimal bandwidth utilization. \textbf{Sparse Workload} applies a Top-K expert selection matrix with column-wise sparsity, where higher sparsity increases load concentration as fewer active experts handle proportionally more traffic. \textbf{Skewed Workload} models Zipfian-driven imbalance in two forms: receiver skew, where many senders target a few experts (incast), and sender skew, where uneven input causes asymmetric expert loads. \textbf{Real MoE Workload} replays traffic traces from MixNet~\cite{mixnet} to reproduce realistic communication patterns. Table~\ref{tab:workloads} summarizes the corresponding distributions.

\textbf{Evaluation Metrics.} We evaluate performance using some key metrics. \textbf{Collective Completion Time (CCT)} measures MoE all-to-all efficiency across average, 80th, 95th, and 99th percentiles, with the 99th percentile approximating total transfer completion. \textbf{Bus Bandwidth (BusBw)} quantifies effective link utilization and enables direct comparison with peak hardware bandwidth to assess algorithmic efficiency. \textbf{NIC Transmission/Reception Volume} captures per-NIC send and receive traffic via a matrix representation, assessing load balance across interfaces. \textbf{Normalized MSE of Load} evaluates intra-domain NIC load balance on a 0–1 scale, where 0 denotes perfect uniformity. \textbf{Iteration Time} measures the total duration of one training iteration.

\textbf{Baseline Methods.} We propose the \textbf{RailS} and compare it with four baseline methods: \textbf{ECMP}~\cite{ecmp} is a widely used equal-cost multipath routing mechanism in datacenters that binds flows to a single path via hashing, providing traditional load balancing and serving as a performance benchmark. \textbf{MinRTT}~\cite{minrtt} is a multipath scheduling algorithm based on RTT measurements. It selects the subflow with the minimum RTT to transmit packets, enabling multipath bandwidth aggregation and load balancing. MinRTT was first implemented in MPTCP. \textbf{PLB}~\cite{plb} is a flowlet-based load balancing scheme that uses IPv6 flow label to dynamically switch paths during idle periods, balancing loads while minimizing packet reordering. \textbf{REPS}~\cite{reps} is a per-packet spraying load balancing algorithm that minimally manages state to reroute packets around congestion hotspots and unreliable links.

\begin{table}[t]
\centering
\caption{The MoE workloads’ token and gating distributions.}
\label{tab:workloads}
\renewcommand{\arraystretch}{1.5} 
\setlength{\tabcolsep}{4.3pt} 
\begin{tabular}{|c|c|c|c|c|c|}
\hline
\diagbox[width=2cm]{Distribution}{Type} & \textbf{Uniform} & \textbf{Sparse} & \makecell{\textbf{Sender-}\\\textbf{skewed}} & \makecell{\textbf{Receiver-}\\\textbf{skewed}} & \makecell{\textbf{Real}\\\textbf{workload}} \\
\hline
\textbf{Token input} & Uniform   & Uniform & Zipf       & Uniform         & \makecell{Uniform} \\
\hline
\textbf{Gating} & Uniform   & Top-K    & Uniform    & Zipf            & \makecell{Training\\-based} \\
\hline
\end{tabular}
\end{table}

\subsection{Performance under Uniform Load}

The uniform-load scenario simulates an ideal condition in which each sender transmits an equal amount of data to all receivers, resulting in balanced link utilization without the presence of ``elephant'' or ``mice'' flows.

\textbf{BusBw.} Fig.~\ref{uniform_busbw} shows the normalized BusBw of different schemes relative to ECMP under the uniform-load scenario. RailS achieves a normalized BusBw of 1.29, representing an improvement of approximately 29\% over ECMP, demonstrating its superior bandwidth utilization. REPS and MinRTT achieve normalized bandwidths of 1.34 and 1.24, respectively, which are close to that of RailS. PLB’s normalized BusBw is 20\% lower than that of RailS, indicating relatively average performance.

\textbf{CDF of CCT.} Fig.~\ref{uniform_fct_cdf} illustrates the CDF of normalized CCT for different schemes relative to RailS under the uniform-load scenario. Specifically, RailS and REPS exhibit similar and concentrated trends, reflecting excellent load-balancing capability. MinRTT completes some flows faster, but its overall completion time is slightly worse than RailS. PLB follows, showing better performance than ECMP.

\textbf{CCT Values.} Fig.~\ref{uniform_fct_perc} further presents the mean, 80th, 95th, and 99th percentile values of normalized CCT for different schemes under uniform load. RailS demonstrates very close mean and tail percentiles, indicating good load-balancing capability. REPS performs consistently with RailS. Although MinRTT achieves the lowest mean CCT, its overall completion time is 4\% worse than RailS. The tail flows of PLB finish relatively simultaneously, with an overall completion time 18\% longer than RailS but still 11\% better than the worst-performing ECMP.

\subsection{Performance under Sparse Load}

The sparse-load scenario is constructed using a Top-K ($K=2$) expert selection matrix to simulate non-uniform communication patterns. By adjusting the sparsity, only a subset of receivers participates in data processing, forming hotspot experts and unbalanced link loads. This setup is used to evaluate the load-balancing capability of different schemes under non-ideal conditions.

\textbf{BusBw.} Fig.~\ref{random_0.4_busbw}, ~\ref{random_0.6_busbw}, ~\ref{random_0.8_busbw} and ~\ref{random_1_busbw} show normalized BusBw of different schemes under various sparsity levels (0.6, 0.4, 0.2, and fully dense). RailS consistently maintains the highest BusBw across all sparsity levels, and its advantage becomes more pronounced as the load becomes sparser. At a sparsity of 0.6, RailS achieves a BusBw $2.8\times$ higher than other schemes. Under other sparsity levels, RailS still achieves 20\%--78\% performance improvements, indicating that RailS fully utilizes the multi-path bandwidth of the Rail architecture to maintain high BusBw.

\begin{figure*}[t]
    \centering
    \begin{minipage}[t]{0.24\linewidth} 
        \centering
        \includegraphics[width=\linewidth]{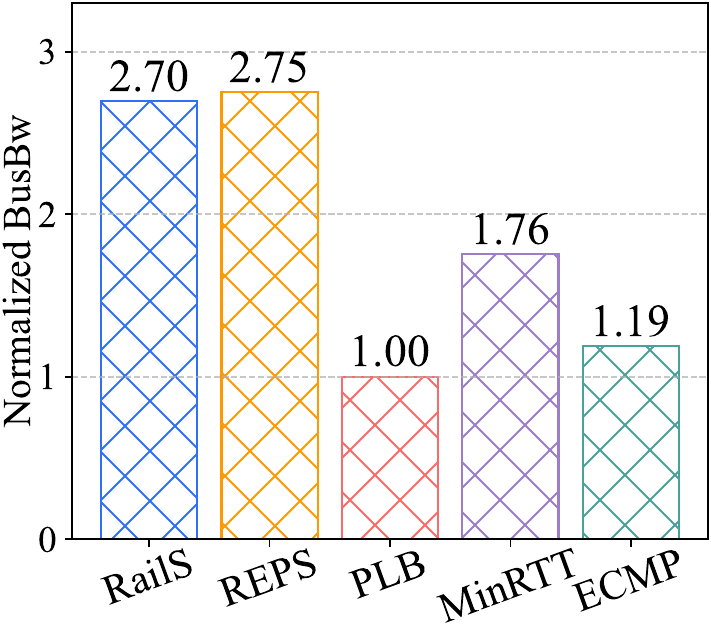}
        \subcaption{Normalized BusBw} 
        \label{sender_skewed_busbw}
    \end{minipage}%
    \hfill
    \begin{minipage}[t]{0.245\linewidth}
        \centering
        \includegraphics[width=\linewidth]{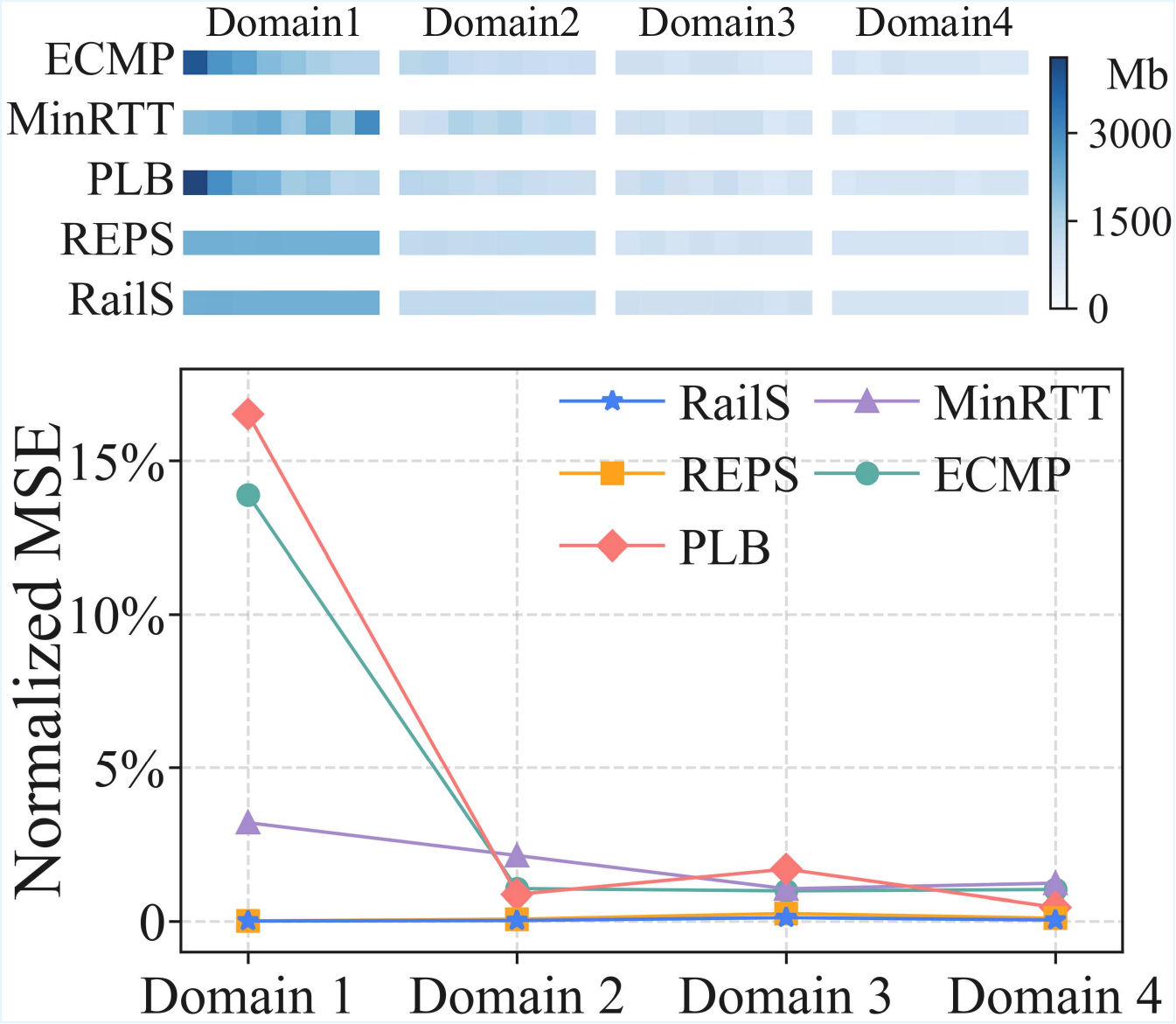}
        \subcaption{Sender load matrix and MSE}
        \label{tx_sender_skewed}
    \end{minipage}%
    \hfill
    \begin{minipage}[t]{0.245\linewidth}
        \centering
        \includegraphics[width=\linewidth]{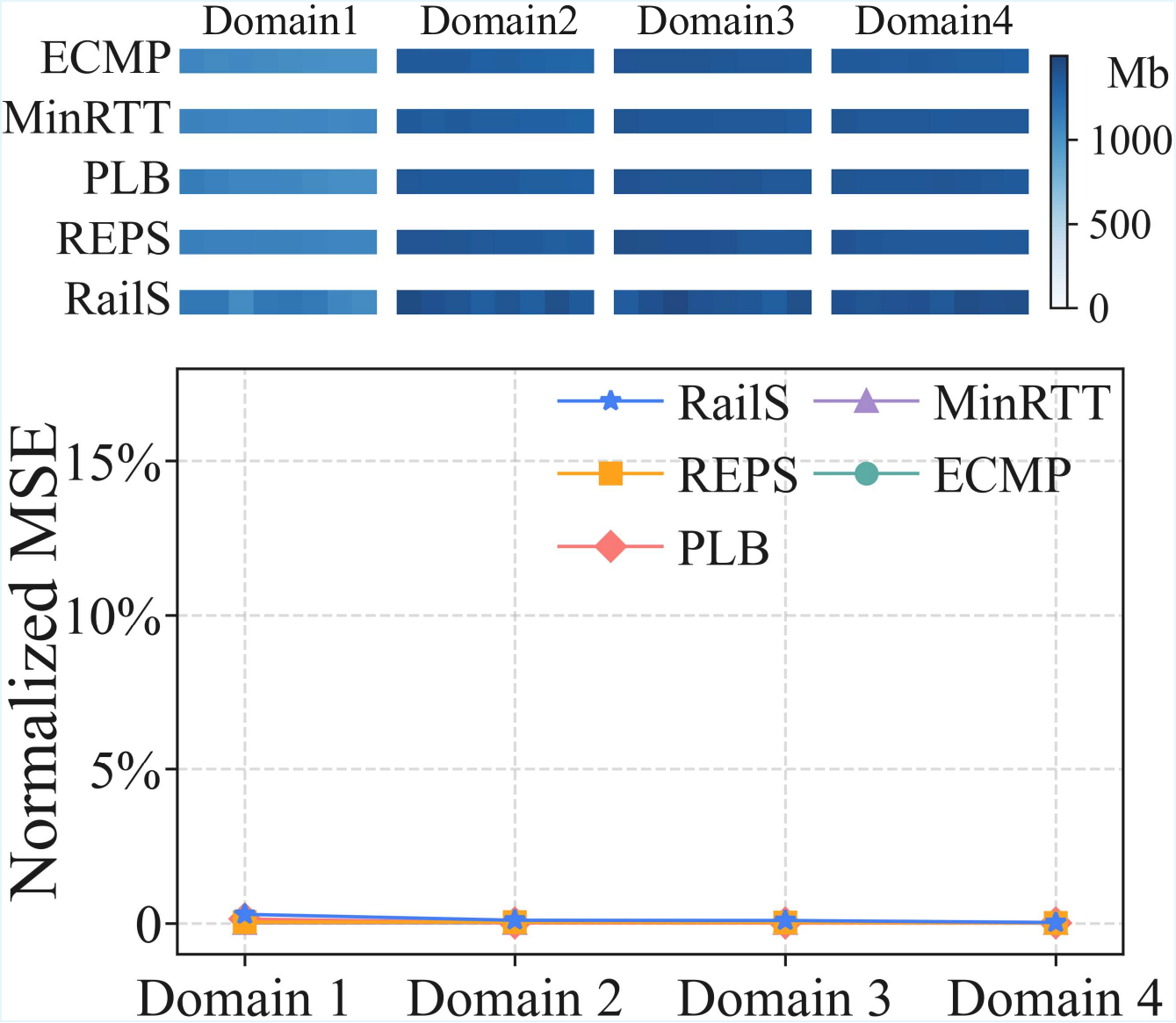}
        \subcaption{Receiver load matrix and MSE}
        \label{rx_sender_skewed}
    \end{minipage}
    \hfill
    \begin{minipage}[t]{0.25\linewidth}
        \centering
        \includegraphics[width=\linewidth]{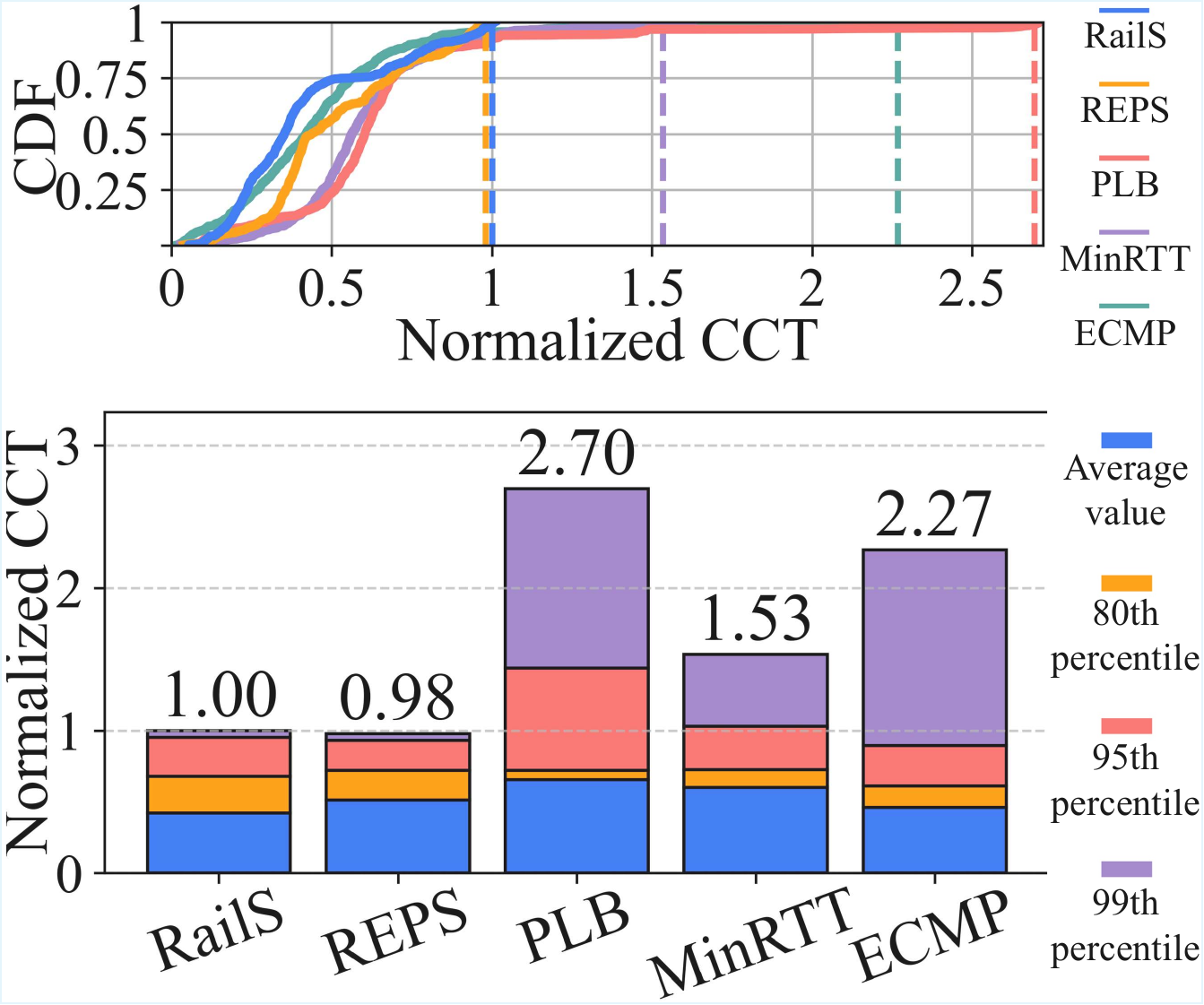}
        \subcaption{Normalized CCT}
        \label{sender_skewed_fct}
    \end{minipage}
    \caption{Performance of different schemes under sender-skewed workloads: (a) normalized BusBw, (b) sender load matrix and its MSE, (c) receiver load matrix and its MSE, and (d) normalized CCT.} 
    
    \label{sender_skewed}
\end{figure*}

\begin{figure*}[t]
    \centering
    \begin{minipage}[t]{0.24\linewidth} 
        \centering
        \includegraphics[width=\linewidth]{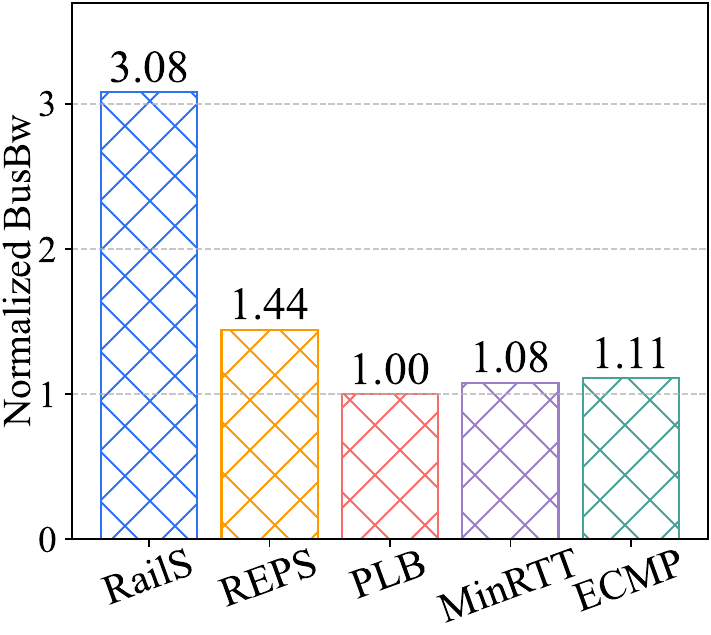}
        \subcaption{Normalized BusBw} 
        \label{receiver_skewed_busbw}
    \end{minipage}%
    \hfill
    \begin{minipage}[t]{0.245\linewidth}
        \centering
        \includegraphics[width=\linewidth]{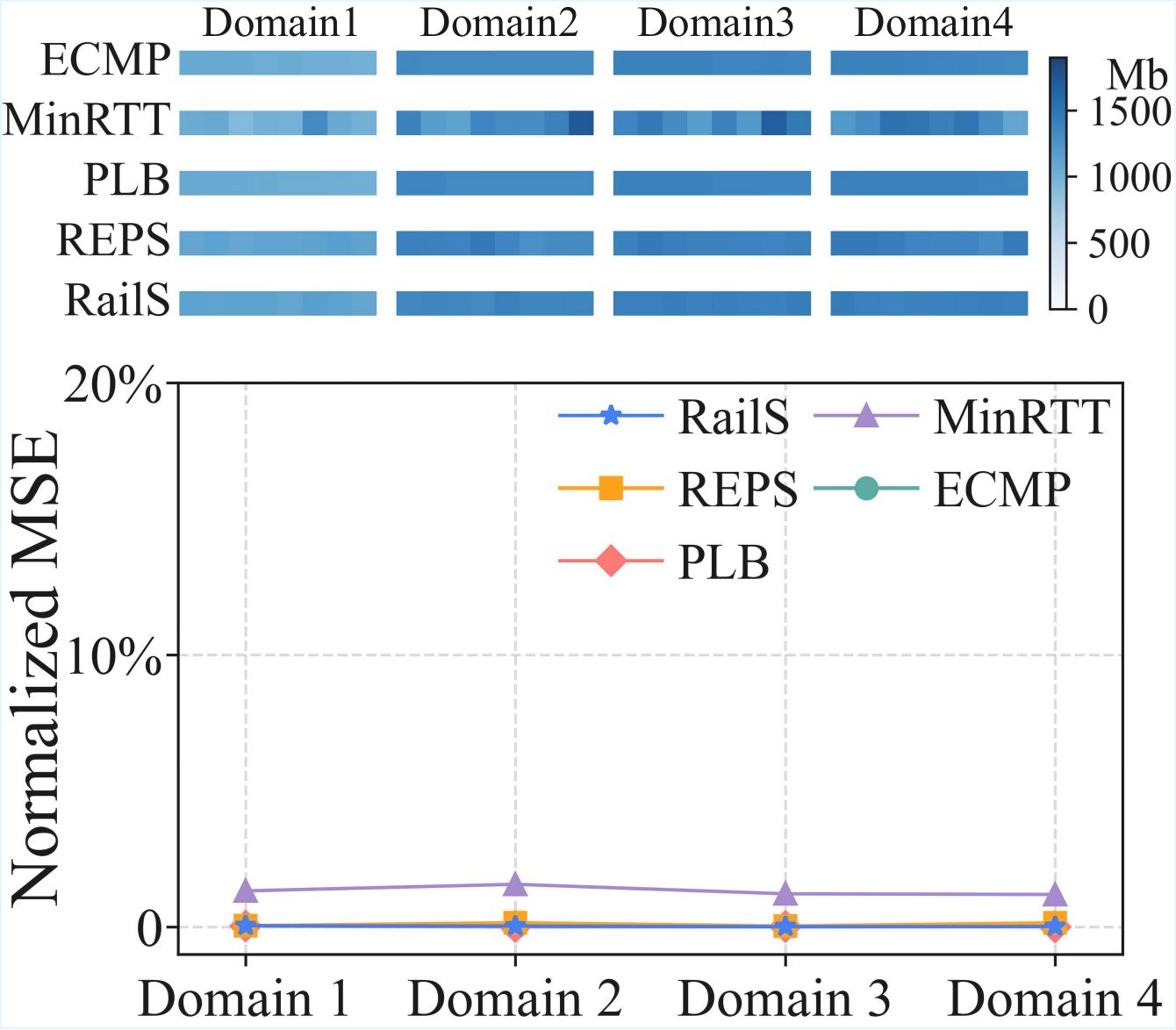}
        \subcaption{Sender load matrix and MSE}
        \label{tx_receiver_skewed}
    \end{minipage}%
    \hfill
    \begin{minipage}[t]{0.245\linewidth}
        \centering
        \includegraphics[width=\linewidth]{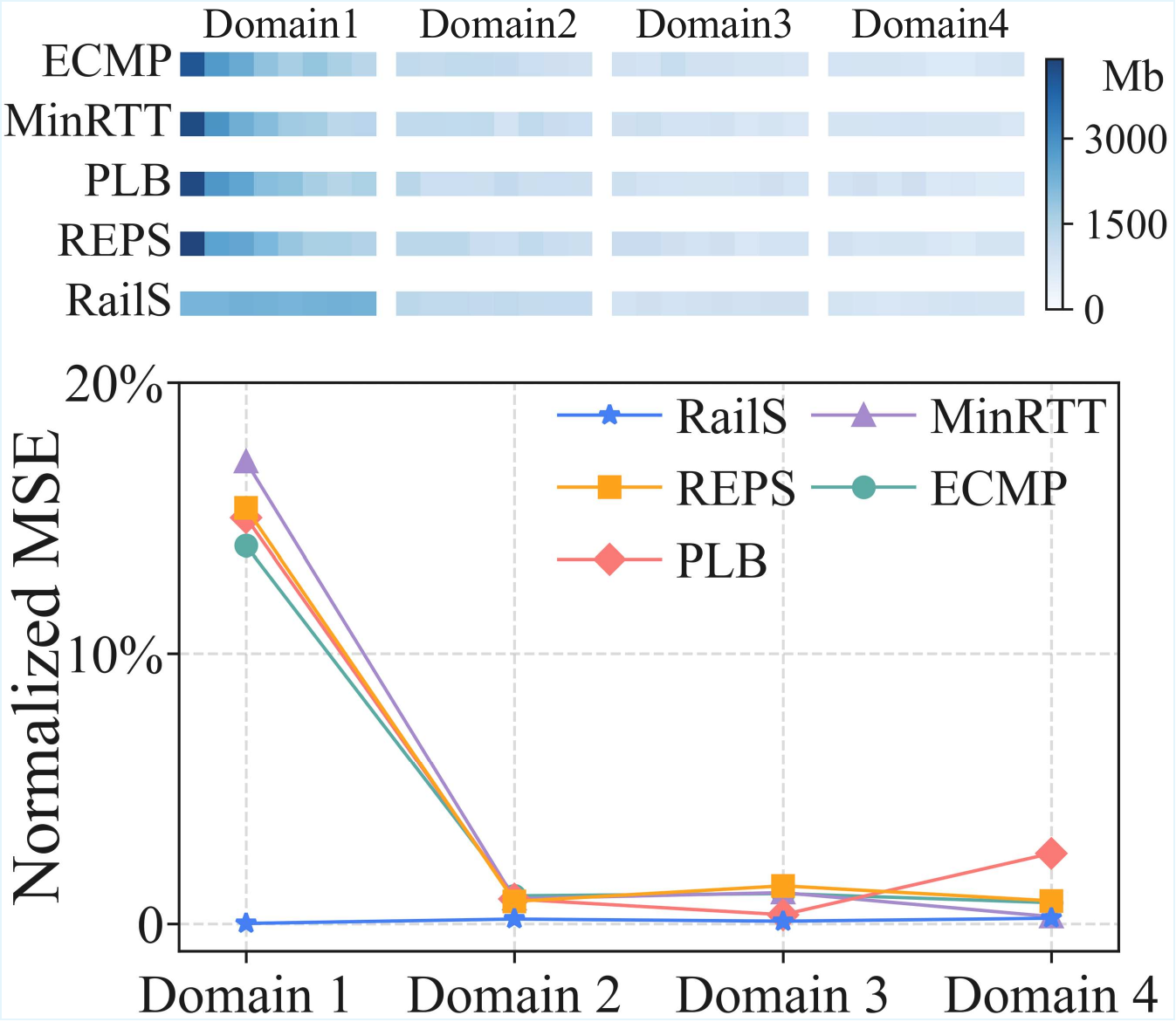}
        \subcaption{Receiver load matrix and MSE}
        \label{rx_receiver_skewed}
    \end{minipage}
    \hfill
    \begin{minipage}[t]{0.25\linewidth}
        \centering
        \includegraphics[width=\linewidth]{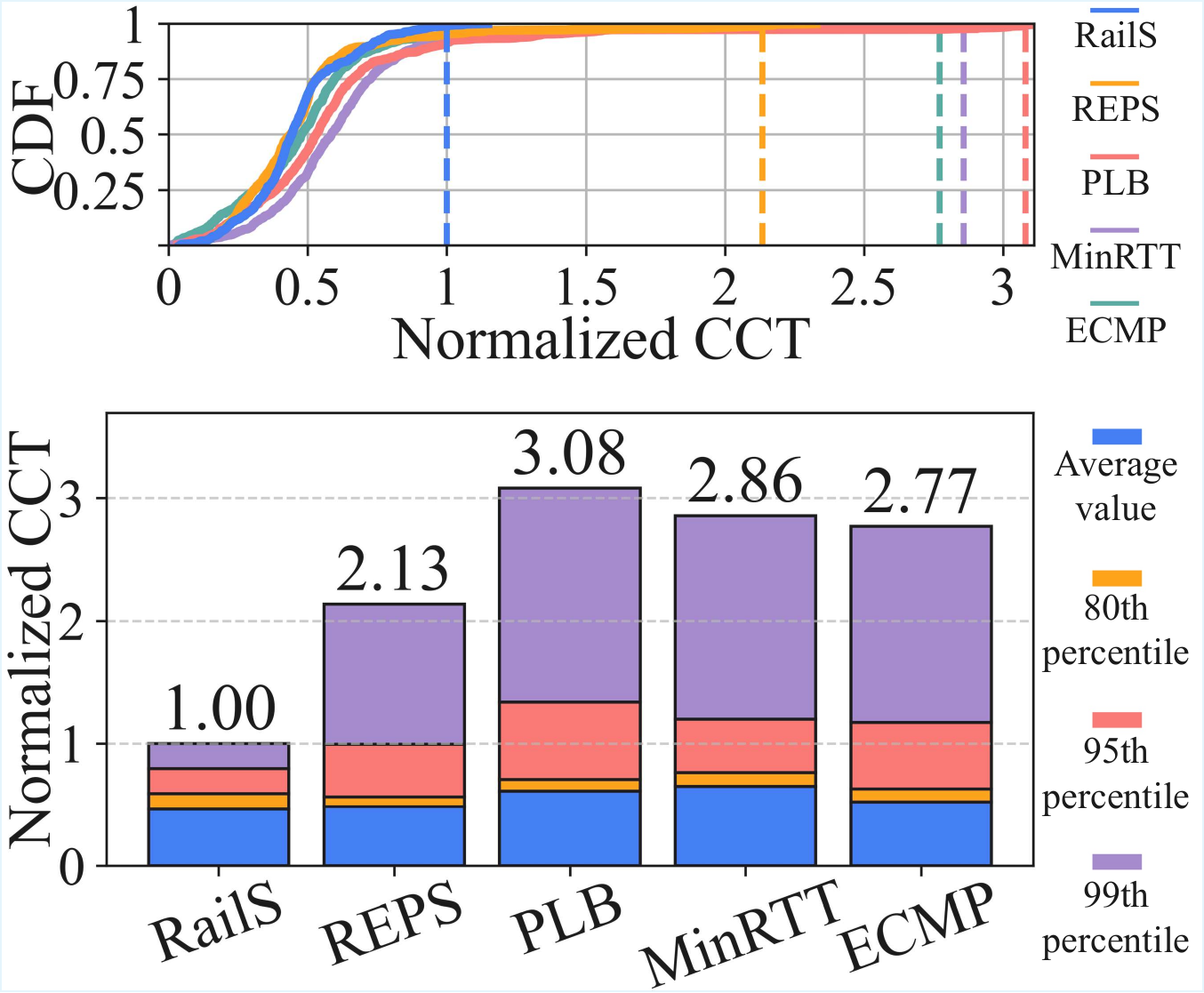}
        \subcaption{Normalized CCT}
        \label{receiver_skewed_fct}
    \end{minipage}
    \caption{Performance of different schemes under receiver-skewed workloads: (a) normalized BusBw, (b) sender load matrix and its MSE, (c) receiver load matrix and its MSE, and (d) normalized CCT.} 
    
    \label{receiver_skewed}
\end{figure*}

\textbf{CDF of CCT.} Fig.~\ref{random_0.4_fct_cdf}, ~\ref{random_0.6_fct_cdf}, ~\ref{random_0.8_fct_cdf} and ~\ref{random_1_fct_cdf} illustrate the CDF of normalized CCT for different schemes under sparse load. It is evident that as sparsity decreases, RailS completes flows faster and with greater advantage. With increasing sparsity, the short-flow performance of RailS becomes closer to other schemes, yet the longest flows still complete the fastest. This demonstrates that RailS achieves optimal load balancing and maintains excellent completion time performance.

\textbf{CCT Values.} Fig.~\ref{random_0.4_fct_perc}, ~\ref{random_0.6_fct_perc}, ~\ref{random_0.8_fct_perc} and ~\ref{random_1_fct_perc} further present the mean, 80th, 95th, and 99th percentile values of normalized CCT for different schemes under sparse load. At low sparsity, RailS maintains closely aligned tail flow completion times. As sparsity increases, the range of tail completion times expands. Compared with other schemes, RailS consistently achieves the best mean and percentile performance in normalized CCT. At a sparsity of 0.6, RailS improves total completion time by more than $1.5\times$. With higher sparsity levels, the improvement remains within 17\%--78\%, demonstrating the superior load-balancing performance of RailS.

\subsection{Performance under Skewed Sender Load}

The sender-skewed scenario employs a Zipf distribution to make a small number of domain experts carry the majority of traffic, forming ``hotspot senders.'' This simulates a situation where non-uniform input leads to extremely unbalanced outgoing loads among experts. It is used to examine each scheme’s bandwidth utilization and congestion mitigation capability under sudden source-side pressure.

\textbf{BusBw.} Fig.~\ref{sender_skewed_busbw} shows that RailS and REPS achieve BusBw $2.7\times$ and $2.75\times$ higher than PLB, respectively. RailS improves bandwidth by 126\% compared with ECMP and by 58\% compared with MinRTT, demonstrating the effectiveness of the spraying mechanism under sender-skewed load.

\textbf{Sender Load Matrix and MSE.} The heatmap in Fig.~\ref{tx_sender_skewed} shows that RailS and REPS maintain balanced traffic distribution across NICs within each domain, with an inter-domain MSE below 0.01, indicating excellent sender load balancing. MinRTT shows moderate balance with an MSE of about 0.03, while PLB and ECMP have the highest MSE, around 15\%.

\textbf{Receiver Load Matrix and MSE.} Fig.~\ref{rx_sender_skewed} shows that although the sending side is skewed, the receiving traffic distribution remains uniform. As a result, the actual NIC-level receiving load of all schemes is balanced, and the bottleneck lies on the sender side.

\textbf{CCT.} As shown in Fig.~\ref{sender_skewed_fct}, the CDF curves indicate that RailS and REPS achieve the best CCT performance, reducing completion time by 55.9\%--62.9\% compared with ECMP and PLB, and by 34.6\% compared with MinRTT.

In summary, RailS and REPS effectively mitigate sender-side bottlenecks under skewed load through the spraying mechanism, achieving balanced transmission. MinRTT benefits from multipath utilization and thus outperforms PLB and ECMP. However, since the bottleneck is not within the network and PLB cannot exploit the Rail architecture’s advantages, its performance is the lowest.

\subsection{Performance under Skewed Receiver Load}

The receiver-skewed scenario models an \textit{incast} situation, where most senders concentrate on a few hot experts following a Zipf distribution, leading to sudden many-to-one traffic bursts. This pattern reflects the real training condition in which ``hot experts'' are frequently invoked. It is used to evaluate each scheme’s ability to suppress tail latency and maintain load balance when the target-side load increases sharply.

\textbf{BusBw.} As shown in Fig.~\ref{receiver_skewed_busbw}, RailS achieves $3.08\times$ normalized BusBw, representing a $2\times$ improvement over ECMP, MinRTT, and PLB, significantly alleviating receiver hotspots. REPS reaches only $1.44\times$ that of PLB, less than half of RailS.

\textbf{Sender Load Matrix and MSE.} Fig.~\ref{tx_receiver_skewed} shows that although the receiver load is skewed, the sending traffic distribution remains uniform. Therefore, all schemes exhibit balanced NIC-level sending and receiving loads, with the bottleneck located on the receiver side.

\textbf{Receiver Load Matrix and MSE.} Fig.~\ref{rx_receiver_skewed} demonstrates that RailS maintains balanced receiving traffic across domains, with an MSE below 0.01, while all other schemes have MSE values around 15\%. The heatmap clearly shows that although Domain~1 experiences a strong receiving skew, RailS effectively evens it out. This indicates that RailS achieves simultaneous load balancing on both sender and receiver sides.

\textbf{CCT.} As shown in Fig.~\ref{receiver_skewed_fct}, RailS achieves the best CCT performance. Compared with other schemes, RailS reduces completion time by more than 50\%, benefiting from balanced NIC utilization on the receiver side, which other schemes fail to achieve.

In summary, although RailS also employs a spraying mechanism, its uniform sending pattern naturally induces uniform receiving distribution, making it significantly superior to REPS. Since the bottleneck lies on the single receiver side rather than within the network, PLB and MinRTT even perform worse than ECMP, as ECMP avoids unnecessary path reorganization in this case.

\begin{figure}[t]
    \centering
    \begin{minipage}[t]{0.49\linewidth} 
        \centering
        \includegraphics[width=\linewidth]{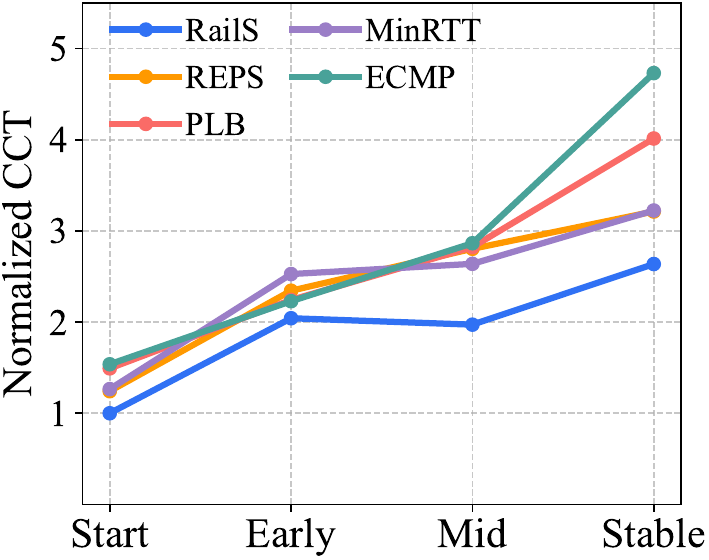}
        \subcaption{Various phases for CCT} 
        \label{full:line}
    \end{minipage}%
    \hfill
    \begin{minipage}[t]{0.49\linewidth}
        \centering
        \includegraphics[width=\linewidth]{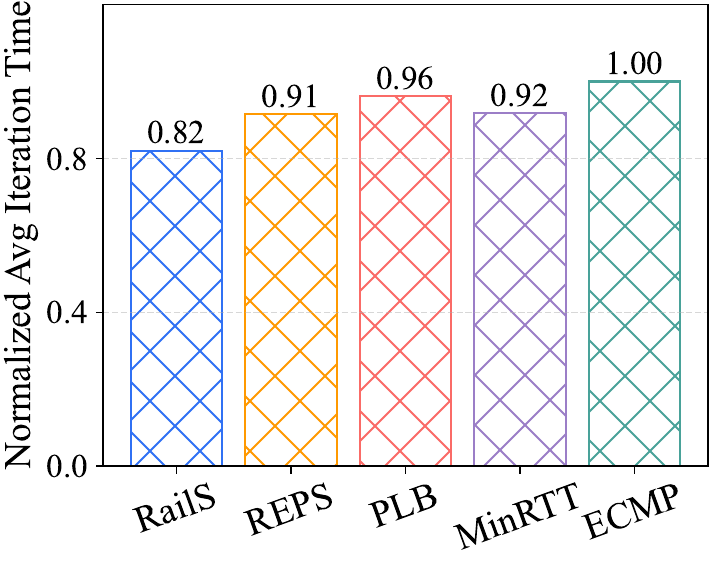}
        \subcaption{Training iteration time}
        \label{full:iteration}
    \end{minipage}%
    \caption{The Mixtral 8×7B model under dense mode exhibits the following metrics: (a) all-to-all completion time and (b) training iteration time.} 
    
    \label{mixtral:full}
\end{figure}

\begin{figure}[t]
    \centering
    \begin{minipage}[t]{0.49\linewidth}
        \centering
        \includegraphics[width=\linewidth]{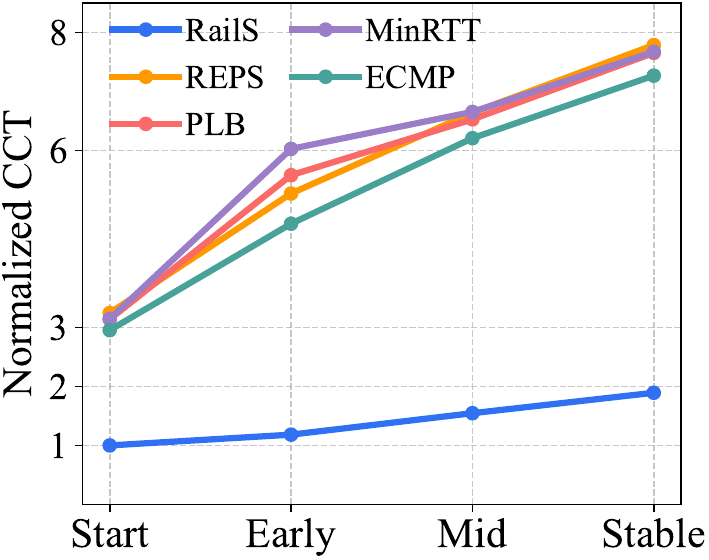}
        \subcaption{Various phases for CCT}
        \label{single:line}
    \end{minipage}
    \hfill
    \begin{minipage}[t]{0.49\linewidth}
        \centering
        \includegraphics[width=\linewidth]{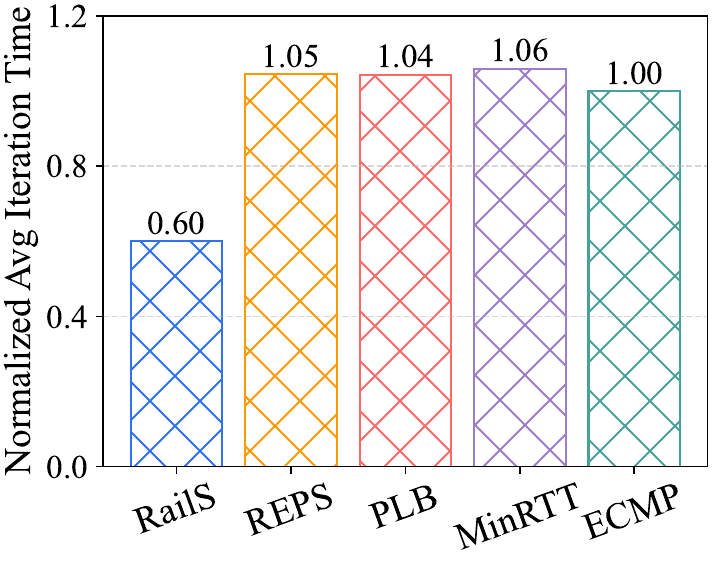}
        \subcaption{Training iteration time}
        \label{single:iteration}
    \end{minipage}
    \caption{The Mixtral 8×7B model under sparse mode exhibits the following metrics: (a) all-to-all completion time and (b) training iteration time.} 
    
    \label{mixtral:single}
\end{figure}

\subsection{Real-World Workload Performance}

We replayed the training and communication emulation of the Mixtral 8$\times$7B MoE training trace. Each expert was deployed across multiple GPUs. The original data to be transmitted by each expert were distributed on GPUs in two different ways to perform all-to-all communication. In the \textbf{dense} setup, each expert’s original data were evenly distributed across multiple GPUs for parallel data exchange. In the \textbf{sparse} setup, the original data were aggregated on a single GPU, and the all-to-all exchange was performed on that GPU.

During training, four phases of the iteration process were identified: \textbf{Start} denotes the first iteration, \textbf{Early} represents the early stage, \textbf{Mid} refers to the middle stage, and \textbf{Stable} indicates the stable stage. The input distribution on each GPU remained uniform, while the total data volume increased as the iteration progressed. For example, in the Start phase, the total transmission volume per expert was around 100~MB, whereas in the Stable phase, it reached 256~MB.

\textbf{Dense Setup.} In Fig.~\ref{mixtral:full}, the normalized CCT analysis indicates that RailS reduces the CCT by 19.5\%--34.9\% in the Start phase, 8.9\%--19.2\% in the Early phase, 25.3\%--31.2\% in the Mid phase, and 17.9\%--44.3\% in the Stable phase. Overall, RailS shortens the iteration time by up to 18\% compared with other schemes.

\textbf{Sparse Setup}. In Fig.~\ref{mixtral:single}, the normalized CCT shows that RailS reduces the CCT by 66.1\%--69.2\% in the Start phase, 75.1\%--80.4\% in the Early phase, 75.1\%--76.8\% in the Mid phase, and 73.9\%--75.7\% in the Stable phase. Throughout the entire training process, RailS achieves at least a 40\% reduction in iteration time compared with other schemes.

These results demonstrate that the sparser the load matrix, the better the performance of the RailS scheme. This is because, under sparse workloads, bottlenecks tend to occur at the receiver side of the Rail architecture rather than within the network. Although some schemes can exploit multipath transmission, they fail to leverage parallel reception, thereby creating receiver-side bottlenecks and performing worse than ECMP. The uniform-sending and uniform-receiving characteristics of RailS fully utilize architectural parallelism and ensure NIC load balancing, achieving optimal performance.

\section{Related Work}
\label{sec7}
The core objective of load balancing is to efficiently distribute network traffic across multiple paths to enhance system throughput and resource utilization. Existing schemes exhibit significant differences in traffic granularity. Traffic granularity can range from individual packets~\cite{reps,handley2017re,uccl_transport,le2024strack} to flowlets~\cite{conga,letflow,expeditus,flowbender}, to subflows~\cite{raiciu2011opportunistic,minrtt,lu2018multi, li2024optimizing} within a single connection, and up to entire connections or flows~\cite{ecmp,wcmp,hedera}. The choice of granularity impacts scheduling accuracy and load balancing, with fine-grained partitioning improving allocation precision and coarse-grained partitioning reducing overhead.

Local load balancing distributes traffic across equivalent paths and can be implemented at the host or switch level. Host-level schemes~\cite{plb,clove,mptcp,zhang2017resilient} offer greater flexibility, easier scaling, and faster adaptation to network changes, supporting multipath scheduling and improving system performance. Switch-level~\cite{pfabric,dcpim,sen2013scalable,microte} schemes react quickly to transient congestion and reduce host overhead but provide limited flexibility for policy customization and scaling.

Traffic forwarding decisions in load balancing rely on network state information. Many schemes utilize connectivity information to identify currently available equivalent paths. Significant differences exist in how load information is applied. Some schemes perform static allocation without relying on load information~\cite{presto,dixit2013impact,olteanu2022edge}. Others dynamically adjust traffic distribution based on local load conditions to optimize link utilization~\cite{detail,kandula2007dynamic,song2023network}. Certain schemes employ global load information for unified scheduling to achieve overall performance optimization~\cite{fastpass,hula}. Recent works further extend load balancing to reconfigurable data center networks, addressing dynamic topologies and time-varying link conditions~\cite{li2022hop, li2024uniform, li2025unlocking}. These differences determine the performance of load balancing schemes in terms of throughput, latency, and network resource utilization.

\section{Conclusion}
\label{sec8}
Driven by the sparse, dynamic, and imbalanced communication of MoE models, we propose RailS, a load balancer for MoE all-to-all communication. RailS exploits the deterministic topology and symmetry of the Rail architecture to resolve path conflicts and structure multi-NIC resource allocation. Theoretically, we prove that sender-side balancing inherently ensures receiver-side balance, reducing global coordination to locally solvable subproblems. System-wise, an LPT-based scheduler efficiently realizes this principle, achieving near-optimal performance through proactive, fine-grained traffic planning. Experiments demonstrate that RailS enhances bus bandwidth, shortens collective communication and training iteration times, and improves overall GPU and network utilization. We hope this work inspires future exploration of topology-aware load balancing and transport co-design, encouraging broader investigation into efficient and scalable communication for large MoE systems.

\bibliographystyle{plain} 
\bibliography{paper}


\end{document}